\documentclass[final,5p,times,twocolumn]{elsarticle}

\usepackage[utf8]{inputenc}
\usepackage{amssymb}
\usepackage{lipsum}
\usepackage{booktabs}
\usepackage{todonotes}
\usepackage{comment}
\usepackage{xspace}
\usepackage{lineno}
\usepackage{xcolor}
\usepackage{nicefrac}
\usepackage{subcaption}
\usepackage{multirow}
\usepackage{amsmath}
\usepackage[normalem]{ulem}
\usepackage[hyphens]{url} 
\usepackage[colorlinks]{hyperref}

\usepackage{graphicx}
\usepackage{fancyhdr}
\usepackage{eso-pic}

\journal{Nuclear Instruments and Methods in Physics Research Section A}

\begin{document}

\begin{frontmatter}

\title{Cherenkov and scintillation light separation in BGO and BSO crystals coupled to SiPMs for dual-readout electromagnetic calorimetry at future colliders}

\author[1,2]{M.~Alviggi}
\author[2,3]{B.~Argiento}
\author[4]{E.~Auffray}
\author[5]{A.~Benaglia}
\author[7,8]{M.~Biasini}
\author[1,2]{V.~Bisignani}
\author[2,6]{D.~Boccanfuso}
\author[2,3]{L.~Borriello}
\author[2]{M.~Campajola\corref{cor1}}
\ead{macampajola@na.infn.it}
\author[7,8]{C.~Cecchi}
\author[1,2]{F.~Cirotto}
\author[2,3]{F.~Conventi}
\author[1,2]{A.~D'Avanzo}
\author[4,9]{J.~Delenne}
\author[1,2]{G.~De Nardo}
\author[1,2]{C.~Di Fraia}
\author[2]{A.~D'Onofrio}
\author[2,6]{L.~Favilla}
\author[2]{M.~Francesconi}
\author[2]{G.~Gaudino}
\author[1,2]{A.~O.~M.~Iorio}
\author[2]{V.~Izzo}
\author[5,11]{M.T.~Lucchini}
\author[8]{E.~Manoni}
\author[2]{M.~Mirra}
\author[8]{S.~Moneta}
\author[2]{P.~Paolucci}
\author[2,3]{S.~Perna}
\author[2]{B.~Rossi}
\author[1,2]{E.~Rossi}
\author[1,2]{J.~Scamardella}
\author[2]{G.~Sekhniaidze}

\cortext[cor1]{Corresponding author}

\address[1]{Dipartimento di Fisica E. Pancini, Universit\`a degli Studi di Napoli Federico II, Naples, Italy}
\address[2]{Istituto Nazionale di Fisica Nucleare, Sezione di Napoli, Naples, Italy}
\address[3]{Università degli Studi di Napoli Parthenope, Naples, Italy}
\address[4]{European Center for the Nuclear Research, Geneva, Switzerland}
\address[5]{Istituto Nazionale di Fisica Nucleare, Sezione di Milano-Bicocca, Milan, Italy}
\address[6]{Scuola Superiore Meridionale, Naples, Italy}
\address[7]{Dipartimento di Fisica e Geologia, Università degli Studi di Perugia, Perugia, Italy}
\address[8]{Istituto Nazionale di Fisica Nucleare, Sezione di Perugia, Perugia, Italy}
\address[9]{Université de Strasbourg, UMR7178 CNRS IPHC, Strasbourg, France}
\address[11]{Università di Milano-Bicocca, Milan, Italy}



\begin{abstract}
We report on the separation of Cherenkov and scintillation light in BGO and BSO crystals read out with silicon photomultipliers~(SiPMs). The two light components are disentangled on an event-by-event basis by combining optical filtering with waveform template fitting, exploiting their distinct spectral and temporal characteristics. Measurements were carried out using high-energy muon and positron beams at the CERN SPS North Area, demonstrating Cherenkov light yields of up to $\sim$150 photoelectrons/GeV in electromagnetic showers.
This work provides the first demonstration of Cherenkov–scintillation light separation in BGO and BSO crystals with SiPM readout, supporting the use of this technology as a building block for a dual-readout electromagnetic calorimeter, as foreseen in the IDEA detector concept for a future $e^+e^-$ Higgs factory.

\end{abstract}

\begin{keyword}
Dual-readout calorimetry\sep Crystals \sep Cherenkov \sep Silicon Photomultiplier (SiPM) \sep Future Circular Collider (FCC)
\end{keyword}

\end{frontmatter}

\tableofcontents 


\section{Introduction}
\label{sec:Introduction}

Future particle physics experiments require significant improvements in detector performance and the development of innovative technologies. A notable example is that of a Higgs factory based on an \( e^+ e^- \) collider, such as the proposed FCC-ee or CEPC projects~\cite{FCC, FCCee, CEPCStudyGroup:2018ghi}. In such facilities, excellent hadronic and jet energy resolution is crucial for reconstructing multi-jet final states, and it therefore constitutes a key aspect in studies of Higgs boson properties and other precision measurements. However, achieving the required level of performance poses significant challenges for the calorimetric system.

Dual-readout calorimetry~\cite{Ferrari:2019vwl,Lee:2017xss} represents a promising approach toward meeting these requirements. By separately measuring the produced scintillation and Cherenkov light, the electromagnetic (EM) fraction of a hadronic shower can be determined  event-by-event, enabling significant improvements in the hadronic energy resolution and response linearity. A widely studied implementation of this concept involves sampling calorimeters that combine scintillating and clear plastic fibers embedded in high-\( Z \) absorbers. This approach has been extensively investigated over more than two decades of  R\&D~\cite{Akchurin:2005eu,Akchurin:2005an,Akchurin:2005rs,Akchurin:2013yaa,Lee:2017shn,Akchurin:2014aoa,Karadzhinova-Ferrer:2022paf,Ampilogov:2023zxb,Albergo:2025exx}, demonstrating the feasibility of achieving single-particle hadronic energy resolution at the level of $\sim$  \( 30\%/\sqrt{E} \). However, the intrinsic sampling fluctuations of these detectors limit their EM resolution to about \( 15\%/\sqrt{E} \), which is significantly worse than what is typically obtained with homogeneous crystal calorimeters.

To simultaneously achieve excellent electromagnetic and hadronic energy resolutions, the IDEA detector concept for FCC-ee~\cite{IDEAStudyGroup:2025gbt} proposes a hybrid calorimetric approach: a segmented homogeneous crystal electromagnetic section followed by a  sampling hadronic calorimeter, both using a dual-readout technique. The electromagnetic section employs high-density scintillating crystals with a total depth of 22 radiation lengths ($X_0$), corresponding to approximately one interaction length ($\lambda_I$), and targets an electromagnetic energy resolution with a stochastic term better than \( 3\%/\sqrt{E} \). 
A high-granularity design is adopted, with a cell transverse size of the order of \( 1 \times 1\,\mathrm{cm}^2 \) and two-layer longitudinal segmentation in order to  enhance soft neutral particle reconstruction and enable high performance of the particle-flow algorithm \cite{Lucchini:2022vss}. 
The inclusion of a traditional electromagnetic homogeneous calorimeter in front of the dual-readout hadronic section would degrade the overall performance of the calorimeter system for shallow-showering hadrons. This limitation can be mitigated by implementing dual-readout capabilities also in the electromagnetic section. Simulation studies within the IDEA design have shown that the target hadronic resolution of $\sim30\%/\sqrt{E}$ can be achieved, provided that at least 50 Cherenkov photoelectrons (\textit{ph.e.}) per GeV of deposited energy are detected in the EM section~\cite{Lucchini:2022vss,Lucchini:2020bac,Lucchini:2022goz}.

Early demonstrations of dual-readout techniques in homogeneous calorimeters, based on coarsely segmented crystals read out with photomultiplier tubes (PMTs), were limited by inefficient light collection~\cite{AKCHURIN2007474,AKCHURIN2008359,AKCHURIN2009710,AKCHURIN201191,Cascella:2013jka}.
Recent developments in crystal manufacturing and photo-detection technologies have paved the way for the development of optimized dual-readout crystal calorimeters. Modern Silicon Photomultipliers (SiPMs) combine compactness, a wide dynamic range, and high photon detection efficiency (PDE) across a broad spectral range~\cite{Garutti:2011qv, Pestotnik:2024amv}, making them particularly suitable for simultaneously detecting scintillation and Cherenkov light. 
New experimental efforts are currently underway to further investigate and optimize dual-readout techniques (\emph{e.g.}~\cite{Hirosky:2024anp, CALA2022166527}), reinforcing the role of crystal calorimeters as a key driver in the development of next-generation experiment detectors.

This work investigates the separation of Cherenkov light in the signals produced by high-energy particles traversing compact geometry homogeneous scintillating crystals of bismuth germanate (Bi\textsubscript{4}Ge\textsubscript{3}O\textsubscript{12}, BGO) and bismuth silicate (Bi\textsubscript{4}Si\textsubscript{3}O\textsubscript{12}, BSO) read out with SiPMs. The method exploits differences in spectral emission, timing, and angular distribution between Cherenkov and scintillation light.
Tests performed with high-energy muon and positron beams at CERN's SPS North Area demonstrate the feasibility of extracting a clear Cherenkov component. These results support the development of dual-readout electromagnetic calorimeters addressing both granularity and readout integration constraints relevant for future \( e^+e^- \) Higgs factories.


\section{Materials and methods}
\label{sec:Materials}
Both BGO and BSO are excellent candidates for electromagnetic calorimetry due to their high density, short radiation length,  small Molière radius ($R_M$), and high scintillation yield. Moreover, their high refractive index enhances Cherenkov photon production~\cite{Cherenkov}. BGO is a well-established material known for its high light yield, while BSO, a relatively newer scintillator~\cite{XIONG2013160}, offers a faster scintillation decay time, although it has a lower light output than BGO. The main properties of the crystals are summarized in Table~\ref{tab:crystal_param}.

\begin{table}[!tbp]
\centering
\caption{Properties of the scintillating crystals under test. Values are taken from~\cite{IDEAStudyGroup:2025gbt} if no superscript is present. Values with superscripts a, b, or c are taken respectively from~\cite{WILLIAMS1996},~\cite{BORTFELD1972}, or~\cite{Benaglia:2025pdf}.}

\vspace{0.2cm}
\begin{tabular}{lcc}
\toprule
\toprule
            	   	           & BGO          & BSO   \\
\midrule
Density [g/cm$^3$]                & 7.1           & 6.8  \\ 
$R_M$ [cm]                        & 2.23          & 2.33 \\ 
$X_0$ [cm]                        & 1.12          & 1.15 \\ 
$\lambda_I$ [cm]                  & 22.7          & 23.4 \\ 
Refractive index at peak emission & 2.15\textsuperscript{a} & 2.06\textsuperscript{b}\\ 
Cherenkov angle ($\theta_C$)      & $62^\circ$ &  $61^\circ$ \\
Scintillation light yield [photons/MeV]            & 3500\textsuperscript{c}    & 950\textsuperscript{c}\\ 
Scintillation emission peak [nm]                & 480          & 470\\
\bottomrule
\bottomrule
\end{tabular} 
\label{tab:crystal_param}
\end{table}

The experimental strategy adopted to isolate the Cherenkov component combines spectral filtering and temporal discrimination.
Cherenkov light is emitted promptly, with a characteristic $\lambda^{-2}$ spectrum and strong directionality, whereas scintillation emission is slower, isotropic, and exhibits a material-dependent emission spectrum and decay time.

\begin{figure} [ht]
\centering
\includegraphics[width=0.99\linewidth]{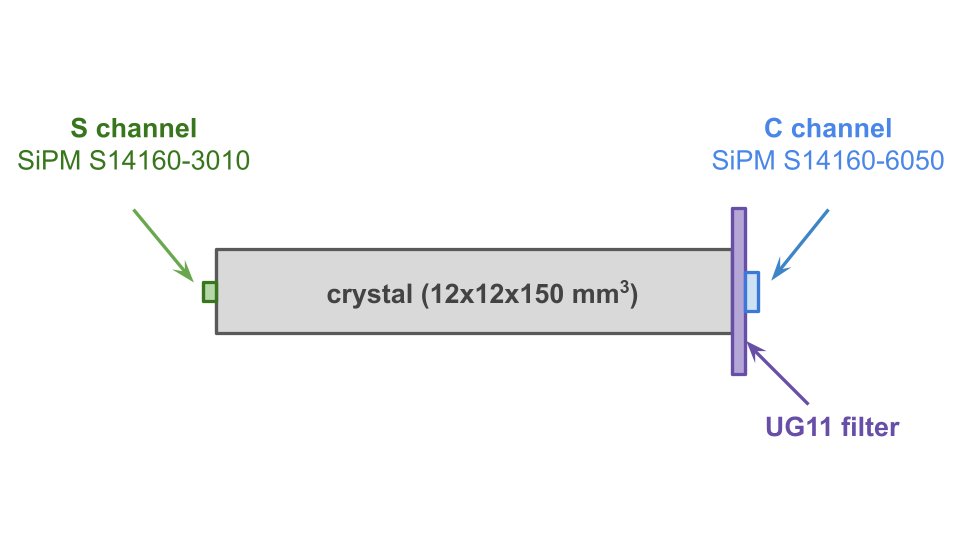}
\caption{Diagram of the  dual-readout scheme using silicon photomultipliers and optical filters. 
}
\label{fig:dr_setup}
\end{figure}

In BGO and BSO crystals, the  scintillation yield significantly exceeds the Cherenkov yield when integrated over the typical sensitivity range of modern SiPMs.
To enable the simultaneous and efficient measurement of both components, each crystal bar is  instrumented with one SiPM on each end face, as shown in Figure~\ref{fig:dr_setup}. 
One readout channel, referred to as \textit{scintillation channel} ($S$), is optimized for detecting  scintillation light, while the other, referred to as \textit{Cherenkov channel} ($C$), is dedicated to the Cherenkov component.
The $C$ channel makes use of a short-pass optical filter placed between the crystals and the SiPM to suppress the sizable longer-wavelength scintillation contribution. 
By acquiring the SiPM waveform and applying a pulse shape analysis, the Cherenkov and scintillation contributions are effectively disentangled, as discussed below.
On the opposite side, the SiPM of the $S$ channel is coupled to the crystal, without any optical filter.

\subsection{Crystal samples}
\label{sec:crystal_samples}
The crystal samples were fabricated by the Shanghai Institute of Ceramics, Chinese Academy of Sciences (SICCAS) \cite{SICCAS}. They are 150 mm long, with a cross-section area of 12~$\times$~12~mm$^2$. Each crystal was wrapped in a 15\,$\mu$m-thick aluminized Mylar foil to prevent light leakage and enhance internal reflection.  The end faces were left unwrapped to allow optical coupling to the photo-sensors. 

Crystal transmittance, scintillation emission spectrum, and decay time are primary factors influencing Cherenkov–scintillation light separation. 

\begin{figure}[h!]
    \centering
        \includegraphics[width=0.99\linewidth]{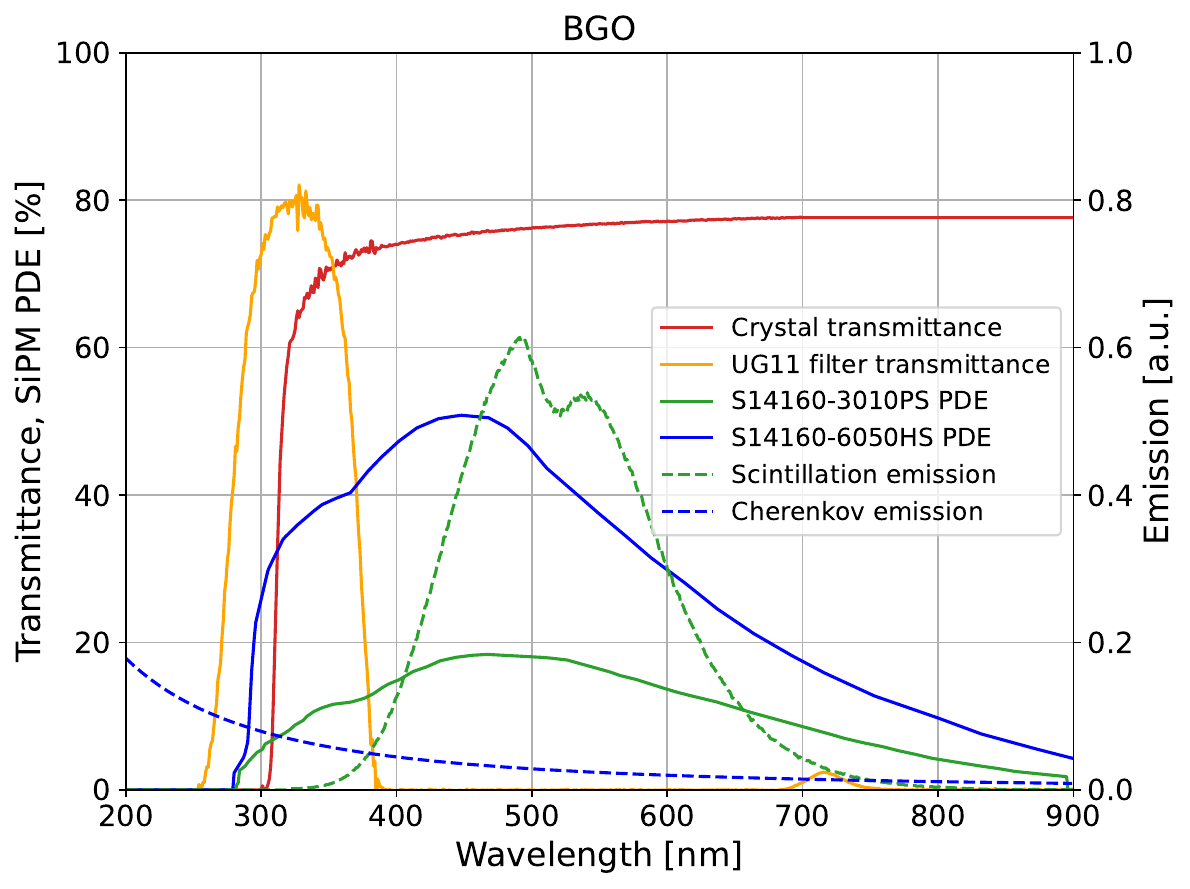}
        \includegraphics[width=0.99\linewidth]{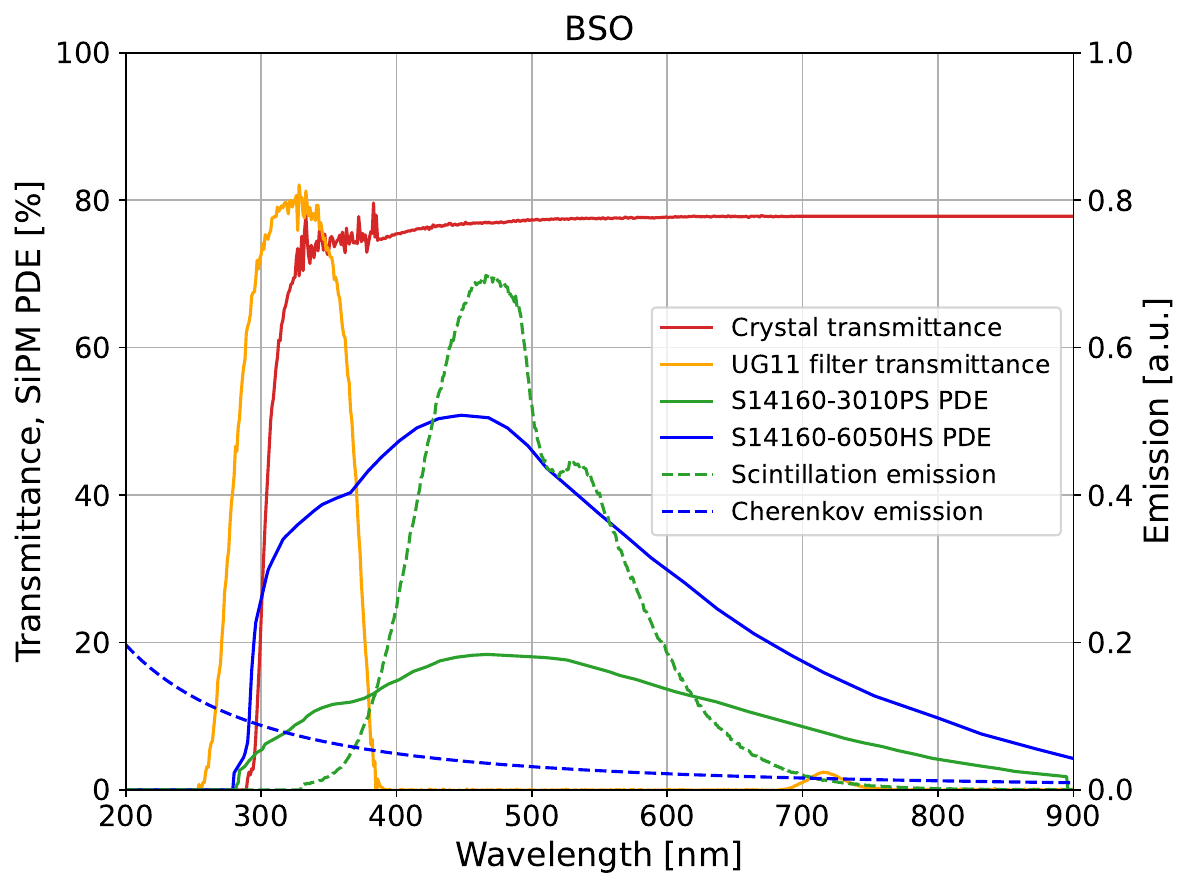}
    \caption{Scintillation and Cherenkov emission spectra, overlaid with the transmission curves of the crystal and the UG11 filter (top figure: BGO; bottom figure: BSO). The oscillations observed in the transmission curves between 300 and 400 nm are instrumental artifacts of the spectrophotometer and do not reflect the intrinsic optical properties of the crystals. The PDE of the $C$ and $S$ channel SiPMs at nominal over-voltage is also shown~\cite{hamamatsu_s14160, hamamatsu_s14160_3010}. }
    \label{fig:emission}
\end{figure}

The crystal transmittance was measured using a Lambda~35~UV-Vis Spectrometer~\cite{LAMBDA25}. 
Results, shown in Figure~\ref{fig:emission}, confirm transparency for photons with wavelengths down to $\sim$300~nm, whereas the scintillation emission peak is at $\sim$480~nm, which matches well with the typical SiPM PDE. 

\begin{figure*}
    \centering
    \includegraphics[width=0.39\linewidth]{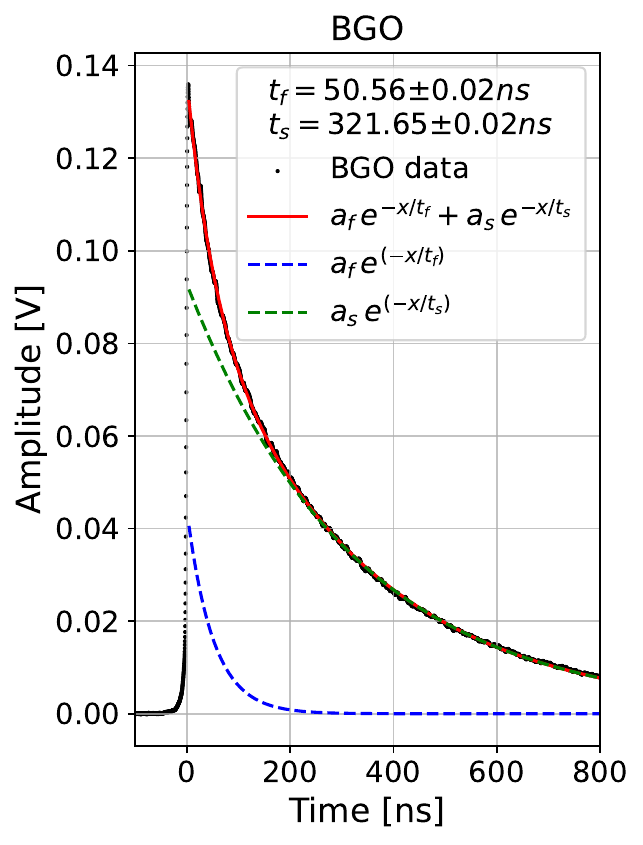}
    \includegraphics[width=0.39\linewidth]{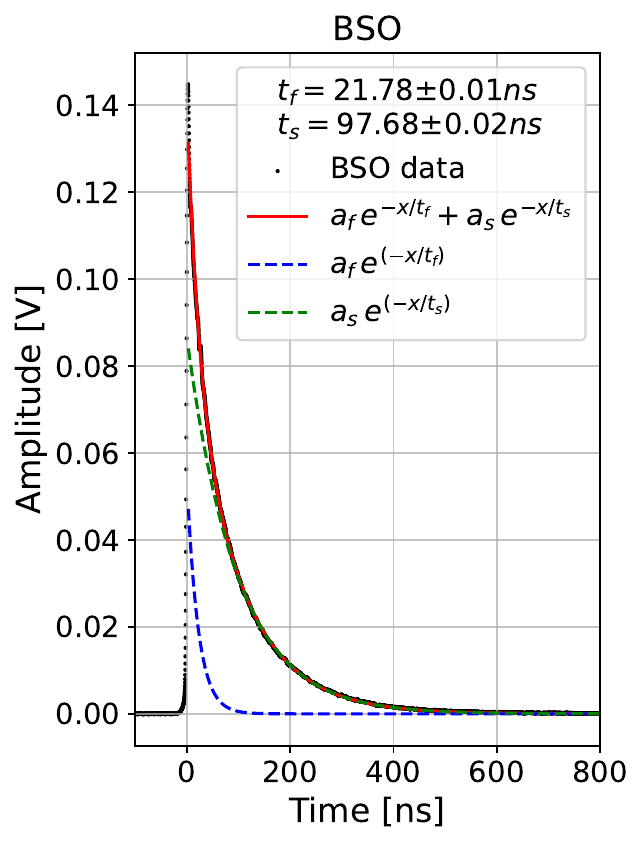}
    \caption{Average waveforms of scintillation light signals measured with a PMT for BGO (left) and BSO (right) crystals under excitation from cosmic rays, shown together with the corresponding best-fit functions. }
    \label{fig:bgo_bso_tau}
\end{figure*}

A characterization of the scintillation decay times at room temperature was performed using a cosmic-ray test bench. Signals were collected with a Hamamatsu R5900 PMT and digitized with the same setup described later in Section~\ref{sec:Test_beam}.
The decay of the scintillation signal is well described by the sum of two exponential components:
\begin{equation}
f(t) = a_f e^{-t/\tau_f} + a_s e^{-t/\tau_s},
\end{equation}
where $\tau_f$ and $\tau_s$ are the decay time constants of the fast and slow components, respectively, and $a_f$, $a_s$ their corresponding relative amplitudes.
The average waveforms and the best-fit model are shown in Figure~\ref{fig:bgo_bso_tau}, while the measured parameters are reported in Table~\ref{tab:bgo_bso_tau}. Decay times of 51~ns and 322~ns for BGO, and 22~ns and 98~ns for BSO fast and slow components, respectively, have been measured. These values are comparable to those reported in the literature \cite{Benaglia:2025pdf}.

\begin{table}[h]
\centering
\caption{Decay time constants ($\tau_f$, $\tau_s$) and relative amplitudes ($a_f$, $a_s$) 
of the fast and slow scintillation components measured for BGO and BSO crystals.}
\begin{tabular}{lcccc}
\toprule
\toprule
Crystal & $\tau_f$ [ns] & $a_f$ [\%] & $\tau_s$ [ns] & $a_s$ [\%] \\
\midrule
BGO & 51 & 32 & 322 & 68 \\
BSO & 22 & 38 & 98 & 62 \\
\bottomrule
\bottomrule
\end{tabular}
\label{tab:bgo_bso_tau}
\end{table}

\subsection{Optical filter}

On the $C$ channel, a short-pass wavelength filter was placed between the crystal and the SiPM to preferentially transmit Cherenkov photons while suppressing most of the scintillation light. The filter used was a SCHOTT UG11 3\,mm-thick colored glass disk with a 1-inch diameter~\cite{Schott:UG11}. 
The filter transmittance was measured with the  spectrometer setup described above and is shown in Figure \ref{fig:emission}. The filter has a narrow bandpass region between 250\,nm~$< \lambda <$~390\,nm, which lies between the BGO and BSO crystals absorption cutoff and the onset of the scintillation emission band, while offering an optical density greater than 15 in the 400\,nm~$< \lambda <$~650\,nm region, where the scintillation emission is significant. 
The residual scintillation light transmitted through the filter is estimated by integrating the scintillation emission spectrum weighted by the UG11 filter transmittance. The transmitted contribution is found to be less than 1\% of the original scintillation signal, arising from the partial transmittance of the filter in the tail of the  emission spectrum, as can be seen in Figure~\ref{fig:emission}.

\subsection{Photodetector readout configuration}
\label{sec:pd_scheme}
The photodetector scheme is optimized for dual-readout operation.
On the $C$ channel, photons transmitted through the filter are detected by a 6\,$\times$\,6\,mm$^2$ Hamamatsu S14160-6050HS SiPM~\cite{hamamatsu_s14160}. It features a 50\,$\mu$m microcell pitch and a total of 14331 microcells. At the nominal bias and at a temperature of 25\,$^\circ$C, the PDE reaches a maximum of $\sim$40\% within the filter bandpass region (Figure~\ref{fig:emission}). 
On the $S$ channel, photons were detected by a 3\,$\times$\,3\,mm$^2$ Hamamatsu S14160-3010PS SiPM \cite{hamamatsu_s14160_3010} with a 10~$\mu$m microcell pitch and 89984 microcells. At the nominal bias and 25\,$^\circ$C, the PDE is $\sim$18\% at 480~nm (Figure~\ref{fig:emission}). 

On the $C$ channel, the large active area and high PDE of the selected SiPM enhance sensitivity within the filter bandpass, improving  the detection of the faint Cherenkov signal.
On the scintillation channel, the small microcell size of the selected SiPM ensures a high dynamic range, suitable for the significantly higher photon density of the scintillation light.

The two SiPMs were mounted on dedicated PCBs equipped with SMA connectors for signal readout. 
The SiPM pairs were operated at bias voltages of 40.7~V ($C$~channel) and 44~V ($S$~channel), corresponding to the nominal overvoltages of 2.7~V and 5~V, respectively.
The bias voltage was provided by a CAEN NDT1419 HV power supply module~\cite{caen_ndt1419}. 
Signals from each SiPM were routed to a CAEN A1423B AC-coupled pre-amplifier~\cite{caen_a1423b}, which provides a selectable gain in the range of 18--54\,dB, a $\pm$1\,V dynamic range, and a bandwidth of 1.5\,GHz. Such a high-bandwidth preamplifier was chosen to preserve the fast temporal structure of the Cherenkov signals on the $C$ channel, enabling pulse-shape analysis for the discrimination between Cherenkov and scintillation components. As an alternative configuration, the SiPMs were read out using a unity-gain circuit with capacitive decoupling and a 50\,$\Omega$ termination to prevent  saturation in the presence of large signals.

\begin{figure}[t]
    \centering
    \includegraphics[width=0.95\linewidth]{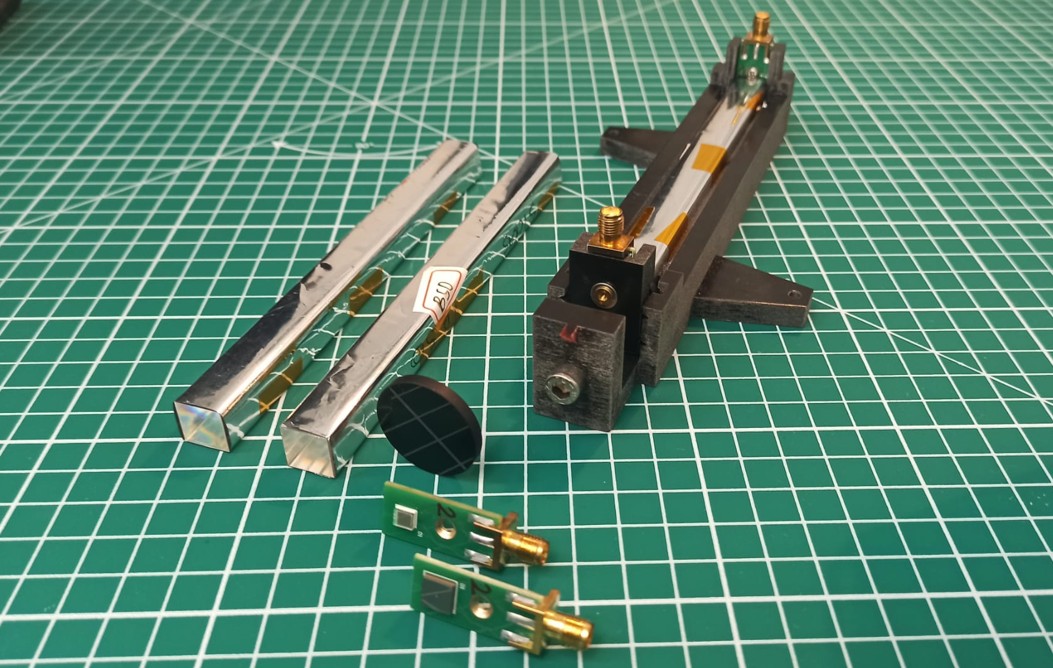}
    \caption{Picture of BGO and BSO crystals wrapped with Mylar, the $C$ and $S$ channel SiPMs, the optical filter, and the crystal holder.}
    \label{fig:pcb_and_holder}
\end{figure}

The crystal, optical filter, and SiPMs were housed in a custom mechanical holder to ensure stable alignment and uniform optical coupling. A thin layer of EJ-550 silicon optical grease~\cite{EJ550}, with a refractive index of 1.46 and a transmittance extending down to approximately 280 nm, was applied at each optical interface for both readout channels to enhance optical coupling.
Figure~\ref{fig:pcb_and_holder} shows  the crystals, optical filter, SiPM boards, and the mechanical holder assembly.

\subsection{SiPM photo-response calibration}
\label{sec:Calibration}
The SiPM photoelectron amplitude calibration was performed on a laboratory bench using a Hamamatsu PLP-10 pulsed laser~\cite{Hamamatsu:PLP10} with an emission wavelength of 405~nm. 
The procedure was carried out using the same readout chain and operating conditions as described in Section~\ref{sec:Test_beam}, with the acquisition triggered on the laser SYNC signal.

\begin{figure*}[t]
  \centering
  \begin{subfigure}{0.44\textwidth}
    \centering
    \includegraphics[width=\textwidth]{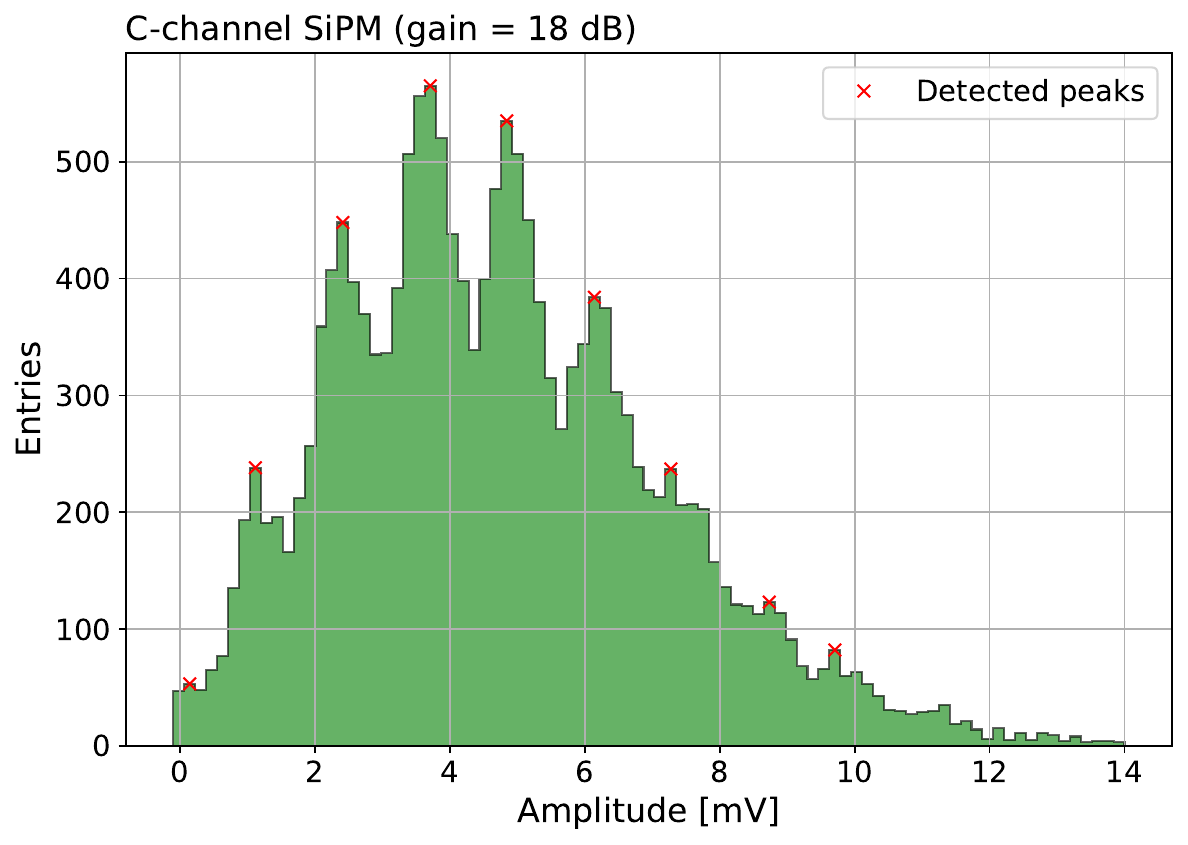}
  \end{subfigure}
  \begin{subfigure}{0.42\textwidth}
    \centering
    \includegraphics[width=\textwidth]{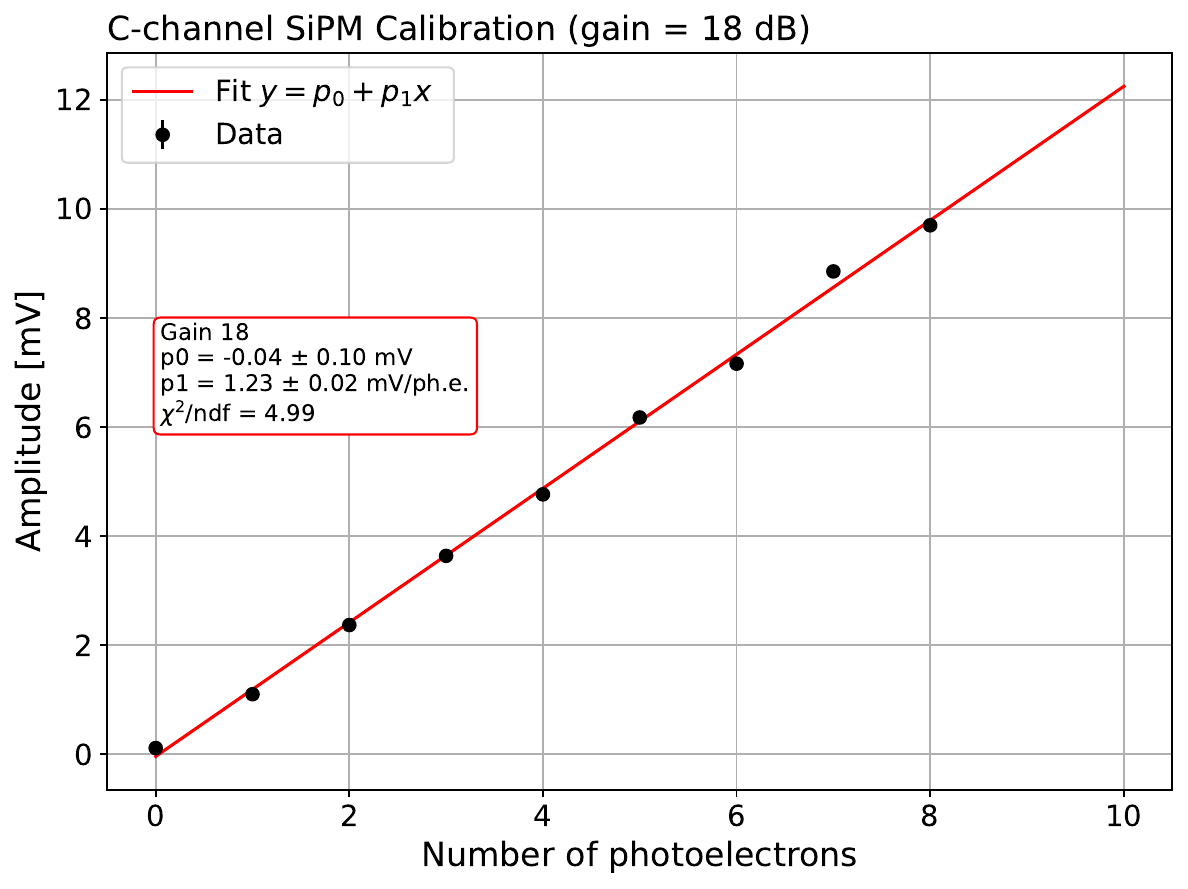}
  \end{subfigure}
  \caption{Left: amplitude spectrum of the $C$ channel SiPM under pulsed light signals with the preamplifier gain set to 18\,dB. Distinct photoelectron peaks are visible, with the positions identified by a peak-finding algorithm and highlighted in red. Right: linear fit (\( y=p_0 + p_1 \cdot x\)) of the $C$ channel SiPM amplitude versus the photoelectron number for a preamplifier gain of 18\,dB.}
  \label{fig:calibration/peakid}
\end{figure*}

For the $C$ channel SiPM, the calibration was performed under attenuated laser conditions to resolve the multi-photon spectrum. Individual photoelectron peaks were identified in the amplitude histogram (Figure~\ref{fig:calibration/peakid} left), and the single photoelectron amplitude was extracted from the slope of a linear fit to the peak positions as a function of the photoelectron number (Figure~\ref{fig:calibration/peakid} right). With the preamplifier gain set to 18~dB, this yielded a value  of ($1.23 \pm 0.02$) mV/$ph.e$.

\begin{figure*}[t]
  \centering
  \begin{subfigure}{0.42\textwidth}
    \centering
    \includegraphics[width=\textwidth]{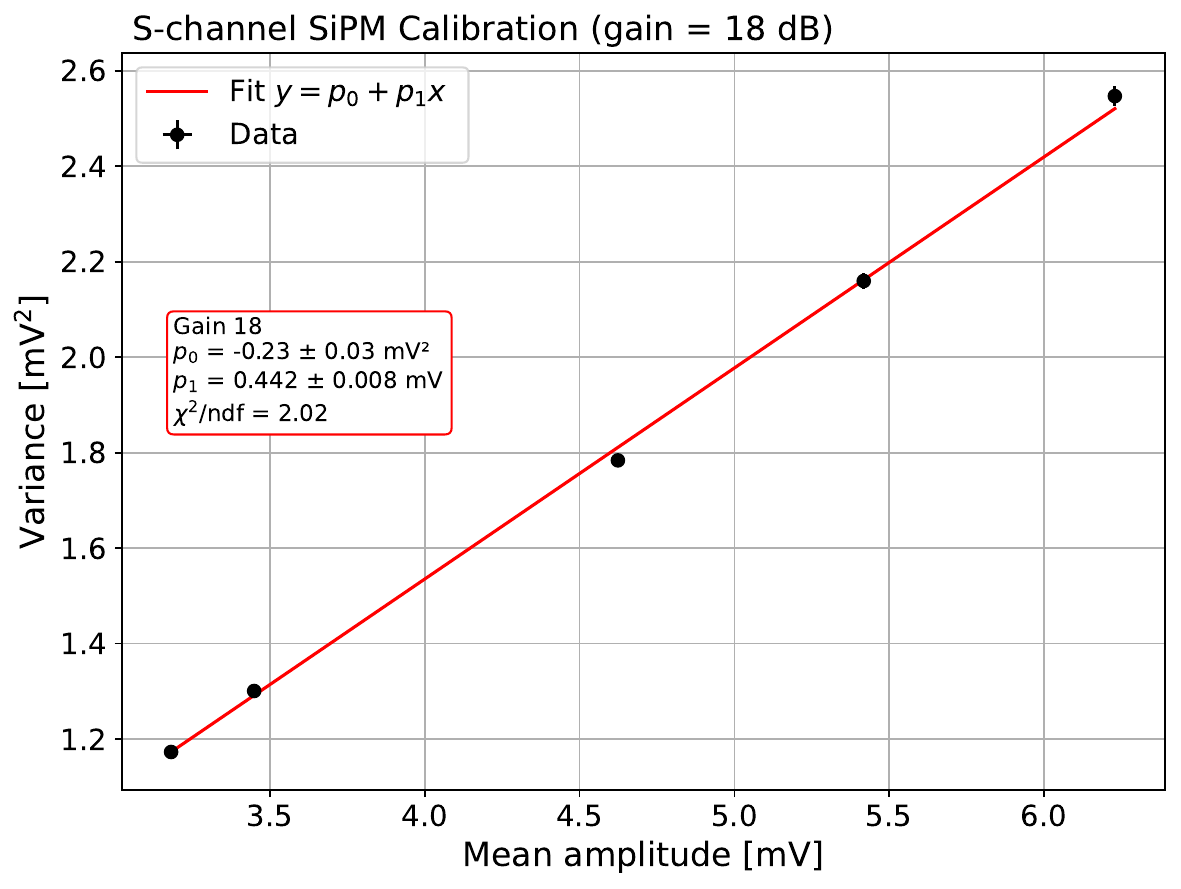}
  \end{subfigure}
  \begin{subfigure}{0.42\textwidth}
    \centering
    \includegraphics[width=\textwidth]{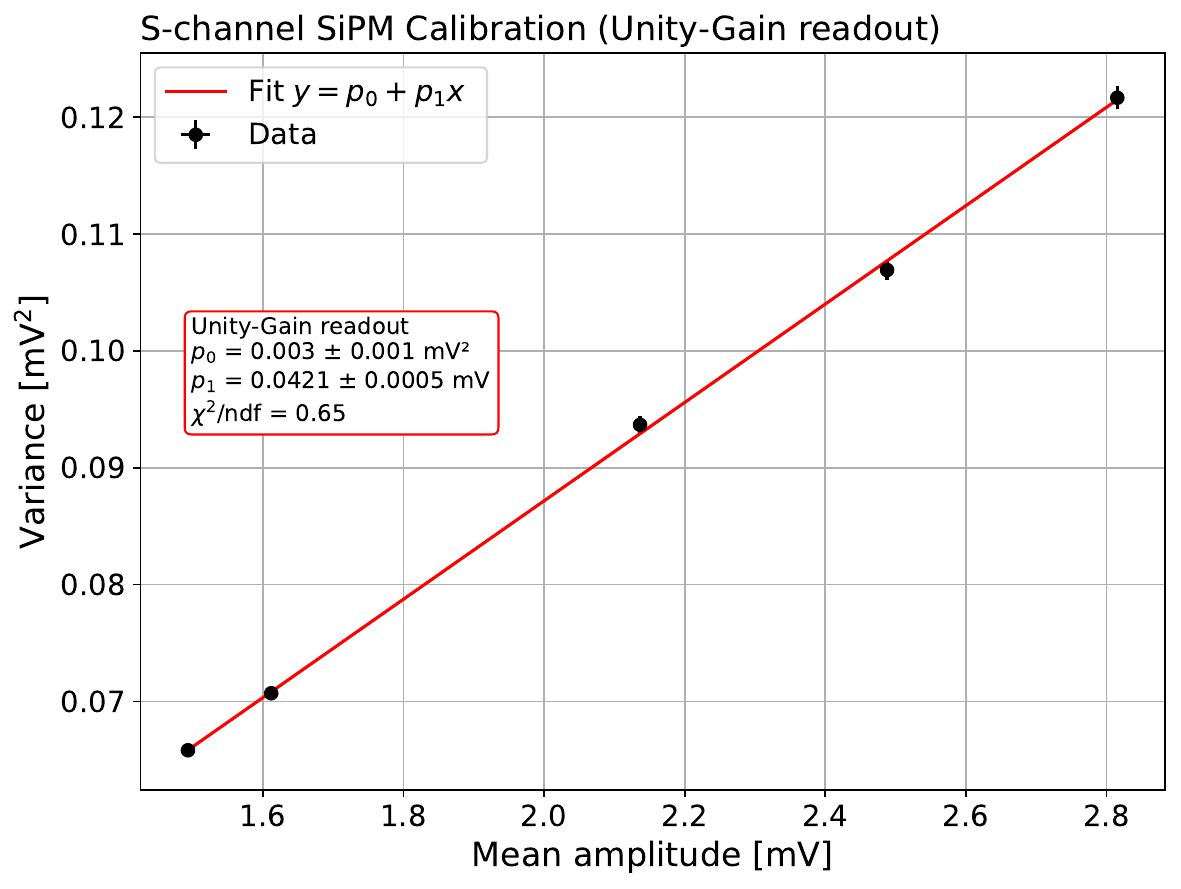}
  \end{subfigure}
  \caption{Linear fit ($y=p_0+p_1\cdot x$) of the variance as a function of the mean signal amplitude at different laser intensities. The slope of the linear fit provides the SiPM calibration factor.  The method applied to the $S$ channel SiPM with a preamplifier gain set to 18 dB (left), and with the unity-gain readout (right).}
  \label{fig:calibration/fitmethodb}
\end{figure*}

For the $S$ channel SiPM, single-photoelectron peaks were not resolved due to limited resolution. In this case, the calibration relied on the statistical properties of the amplitude distribution: assuming Poissonian photon statistics, the conversion factor was extracted from the slope of a linear fit to the variance as a function of the mean signal amplitude at different laser intensities, as shown in Figure~\ref{fig:calibration/fitmethodb}. With the preamplifier gain set to 18~dB, this corresponds to a calibration factor of ($0.442\pm 0.008$)\,mV/$ph.e$, while using the unity-gain circuit it was ($0.0421\pm 0.0005$)\,mV/$ph.e$.
The consistency of the two gain extraction methods was validated by applying the second method to the $C$ channel SiPM, yielding results that agree within 12\% with those obtained from the photoelectron peak method. This difference was taken as an additional systematic uncertainty on these calibration factors.

A charge calibration was derived from the amplitude calibration using a conversion factor defined as the ratio of the signal waveform integral to its amplitude, evaluated on an average laser waveform. 
This procedure provided calibration factors of ($192\pm 19$)\,mV\,ns$/ph.e.$ for the $C$ channel at a gain of 18~dB, while $(23\pm 2$)\,mV\,ns$/ph.e.$, and ($1.8\pm 0.2$)\,mV\,ns$/ph.e.$ on the $S$ channel at a gain of 18~dB and using the unity-gain readout, respectively.


\section{Test beam}
\label{sec:Test_beam}

Both BGO and BSO crystals were tested at the CERN SPS North Area (H6 beam line~\cite{H6}) using high-energy  muon and positron beams. 

\subsection{Setup description}

\begin{figure*}[t!]
\centering
    \includegraphics[width=0.8\linewidth]{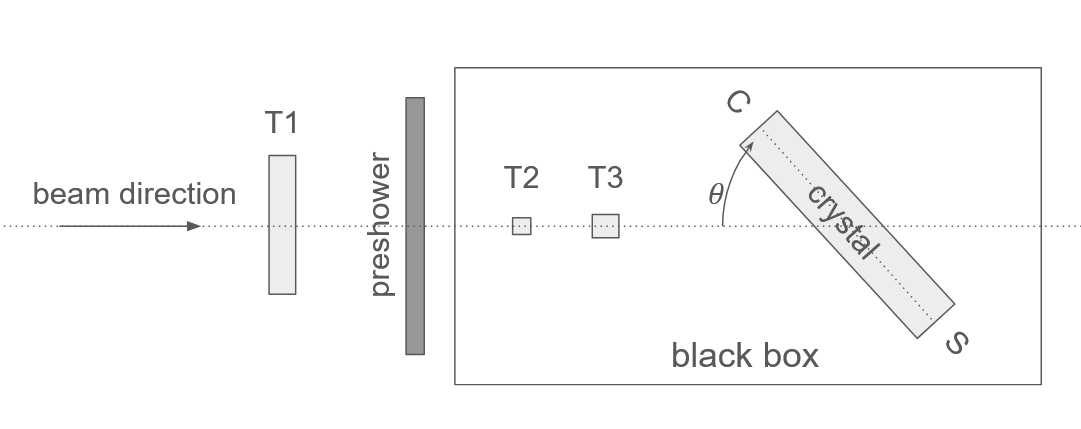}
\caption{Schematic of the experimental setup during the test beam. The figure shows the crystal under test instrumented with the two readout channels, $C$ and $S$, and the three scintillators T1, T2, and T3 used to form the trigger decision. Object sizes and distances are not to scale.}
\label{fig:tb_setup}
\end{figure*}

A schematic of the experimental setup used during the test beam is shown in Figure~\ref{fig:tb_setup}. 
Each crystal, housed together with the SiPMs and the optical filter, was mounted horizontally on a remotely controlled rotating stage, allowing rotation around the vertical axis. 
The rotation angle $\theta$ is defined so that at $\theta=0^\circ$ the beam is aligned with the crystal longitudinal axis, impinging first from the $C$ channel side. Varying $\theta$ allowed an angular scan of the crystal axis with respect to the incident beam direction.

The SiPM pairs were operated at nominal bias and read out through the CAEN A1423B preamplifier with a gain of 18~dB. Only during electron beam runs with BGO crystals, where the scintillation yield was particularly high, the $S$ channel was instead read out using a unity-gain circuit to avoid signal saturation.

Three scintillator units, named T1, T2, and T3, were placed along the beam path upstream of the crystal to provide a trigger signal to the data acquisition (DAQ) system. 
T1 and T3 were plastic scintillators with dimensions of $5\times5\times1$~cm$^3$ and $1\times~1\times1$~cm$^3$, respectively, while T2 was a 3 \(\times\) 3 \(\times\) 3 mm$^3$ LYSO crystal. T2 and T3 were mechanically aligned with the center of the crystal under test. T1 was read out by a PMT, while T2 and T3 by two SiPMs. Depending on the run type, the trigger signal was generated by the coincidence of T1 with either T2 or T3 to geometrically select particles impinging on the crystal.
A 1~cm--thick lead tile ($\approx 1.8\,X_0$), acting as a passive preshower, was placed between T1 and T2 to ensure shower development during $e^+$ runs and thus a sizable energy deposit, particularly for 90$^\circ$ crystal orientation with respect to the beam.

\begin{figure}[t]
\centering
    \includegraphics[width=0.8\linewidth]{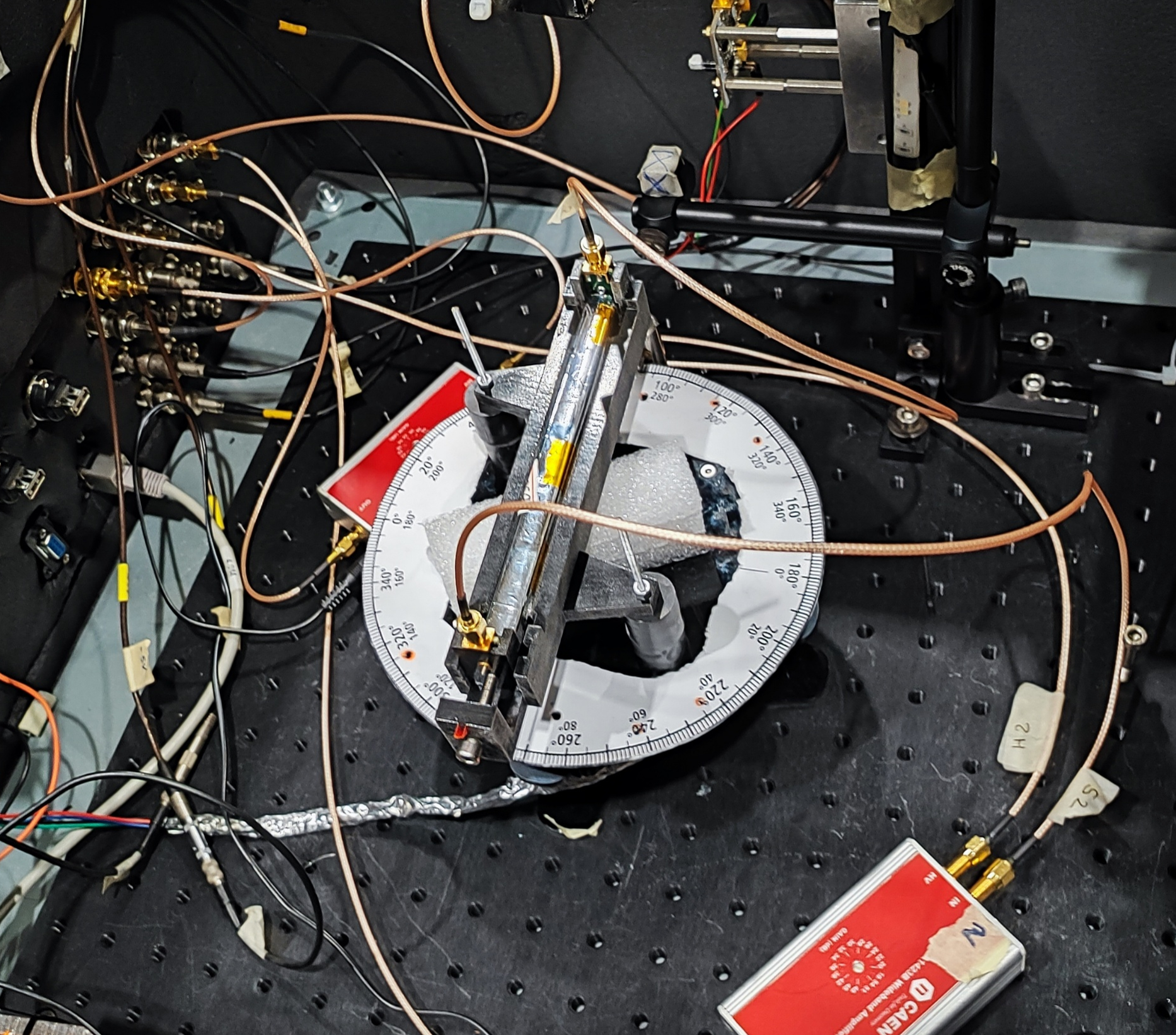}
\caption{Picture of the experimental setup for the test beam, showing the crystal and the readout SiPMs mounted on the rotator stage inside the dark box.}
\label{fig:tb_setup_photo}
\end{figure}

The crystal setup, along with the SiPM preamplifiers and the T2 and T3 scintillators, was enclosed in a steel box (Figure~\ref{fig:tb_setup_photo}) located along the beam line. The thickness of the box wall corresponds to approximately 0.085$X_0$.
The box served as a light-tight enclosure and a Faraday cage to protect against electromagnetic noise. The box was thermally insulated from the environment by a 1 cm--thick expanded nitrile sponge foil. Inside the box, a copper radiator connected to an external chiller via a closed-loop water-glycol circuit, equipped with a remote-controlled fan, maintained a stable operating temperature of 23$^\circ$C. The temperature inside the box was monitored using a thermometer connected to a micro-controller board, which implemented a Proportional-Integral-Derivative (PID) feedback system to control the chiller's target temperature.

The box was placed on a three-axis moving table approximately 3 meters downstream from the beamline exit window. 

All signals from the trigger scintillators and the crystal SiPMs were digitized using a Tektronix MSO46B 6 channel oscilloscope~\cite{tektronix_mso46b} with a 6.25 Gs/s sampling rate, 1.5~GHz bandwidth, and 12--bit ADC. 
The DAQ system included a front-end computer in the experimental hall, responsible for reading continuously digitized waveforms from the oscilloscope, managing slow control from the temperature sensor and SiPM power supplies, storing data, and controlling the run. 
The DAQ software was developed using the MIDAS framework~\cite{MIDAS}.

\subsection{Run configurations}

The setup was tested using beams of $\mu^+$ at 120~GeV and $e^+$ at 10~GeV. This allowed studying the crystal response to minimum ionizing particles (MIPs) and electromagnetic showers.  
For each crystal and particle species, several data-taking runs were conducted for different  orientations of the crystal with respect to the beam. This enabled the investigation of the scintillation response as a function of different energy depositions, as well as the study of the Cherenkov and scintillation signal separation as a function of the beam incident angle.

The beam intensity was approximately $10^4$ particles per spill for electrons and  about one order of magnitude lower for muons, with a typical spill duration ranging between 4.8 s and 9.6 s. Under these conditions, no pileup effects were observed.

Each run file contained between 5000 and 40000 events, depending on the configuration, along with an automatic dump of all the HV settings, slow control, and readout parameters at the beginning of the run. 

The acquisition window for each event was set 2~$\mu$s for runs with BGO and 1~$\mu$s for runs with BSO.


\section{Results}
\label{sec:results}

\subsection{Scintillation light output and energy response} \label{sec:results_schannel}

Runs with muons were used to evaluate the relation between the $S$ channel  light output and the energy deposited in the crystal. 
Since the energy loss of a MIP is approximately proportional to its path length through the material, the angular scan runs were used to probe different effective thicknesses and thus a range of energy deposit values. To support the interpretation of the experimental results, we performed a simulation of the detector setup using \textsc{Geant4} (version 11.3.2)~\cite{GEANT4}. The goal was to accurately determine the energy deposited by MIPs in the crystal to provide a reliable normalization of the measured light output.

\begin{figure}[htp!]
    \centering
    \includegraphics[width=0.99\linewidth]{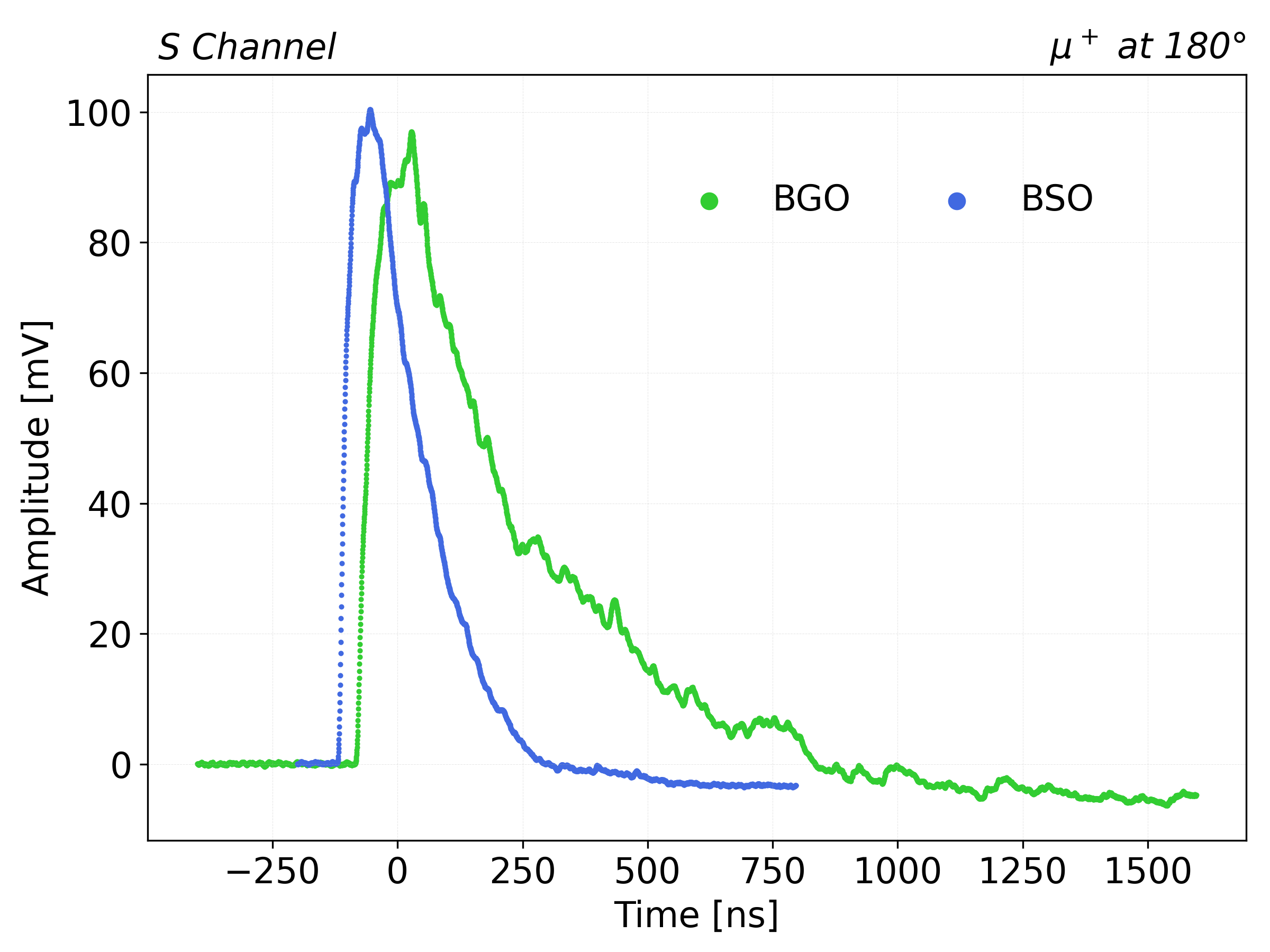}
    \caption{Example of waveforms corresponding to the passage of a MIP, recorded on the $S$ channel for BGO and BSO crystals at $180^\circ$ (crystal axis aligned with the beam direction) in $\mu^+$ beam runs.}
    \label{fig:mu_Schannel_waveforms}
\end{figure}

\begin{figure*}[ht]
    \centering
    \begin{subfigure}{0.43\linewidth}
        \includegraphics[width=1\linewidth]{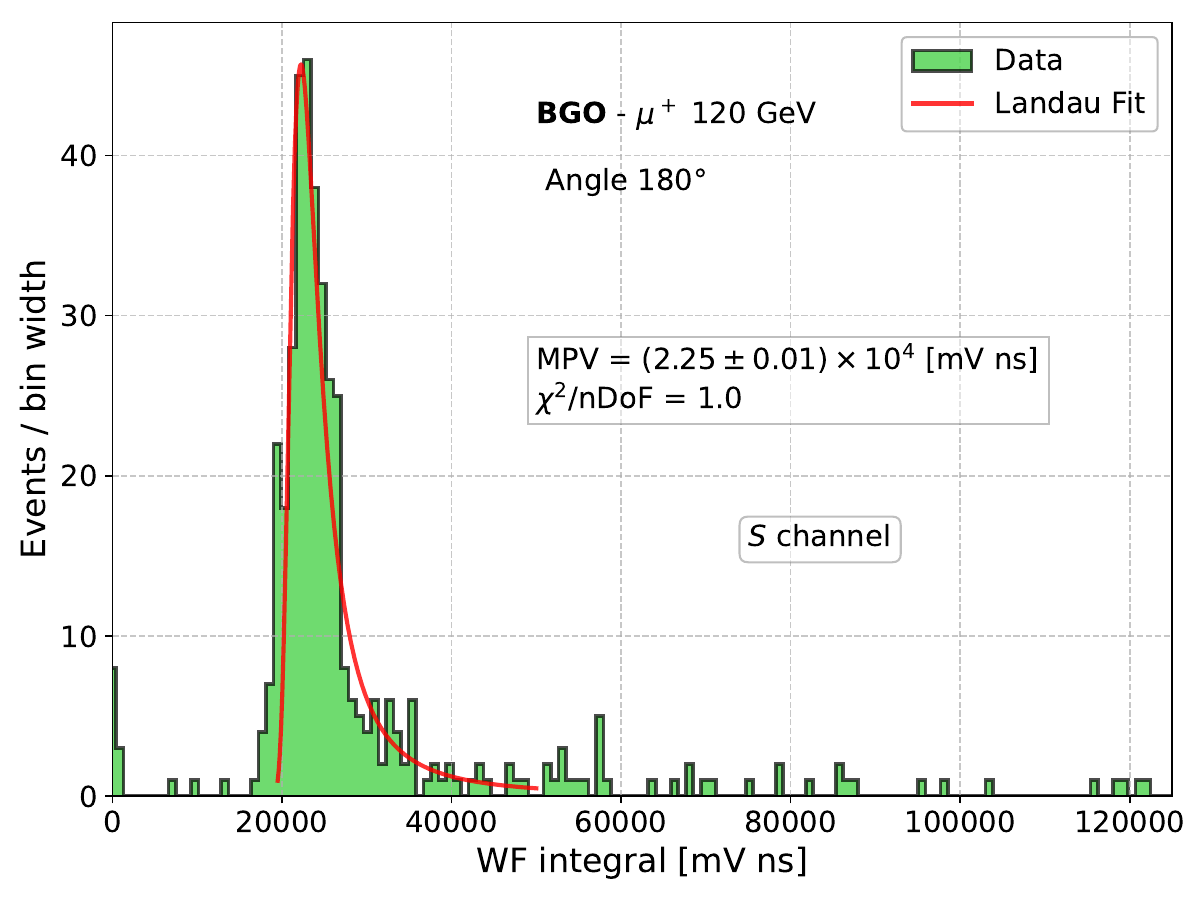}
        \end{subfigure}
    \begin{subfigure}{0.43\linewidth}
        \includegraphics[width=1\linewidth]{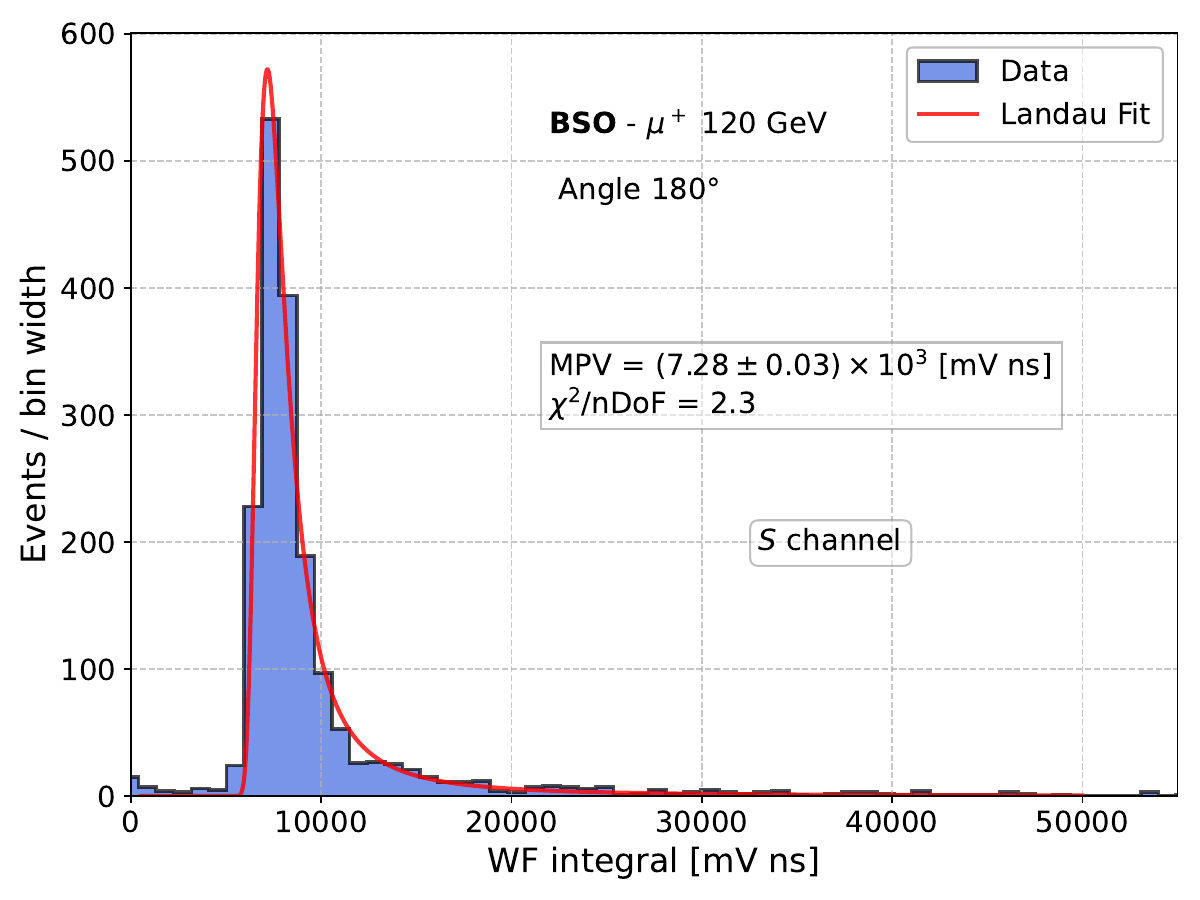}
        \end{subfigure}
    \caption{Example of Landau fit to the waveform integral distribution. This plot refers to $S$ channel signals of BGO (left) and BSO (right) crystal at $180^\circ$ in $\mu^+$ beam runs.}
    \label{fig:landau_fit}
\end{figure*}

\begin{figure*}[ht]
    \centering
    \begin{subfigure}{0.45\linewidth}
    \includegraphics[width=1\linewidth]{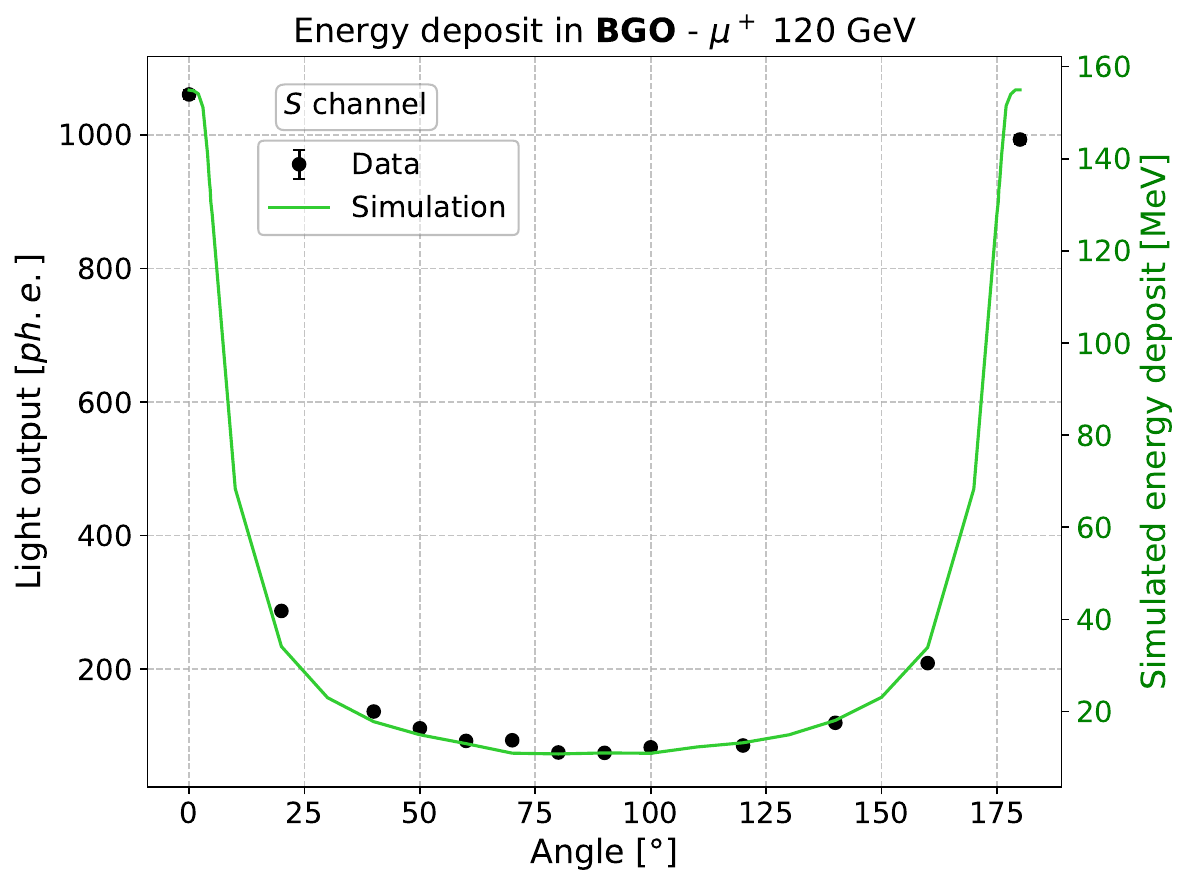}
    \end{subfigure}
    \begin{subfigure}{0.45\linewidth}
    \includegraphics[width=1\linewidth]{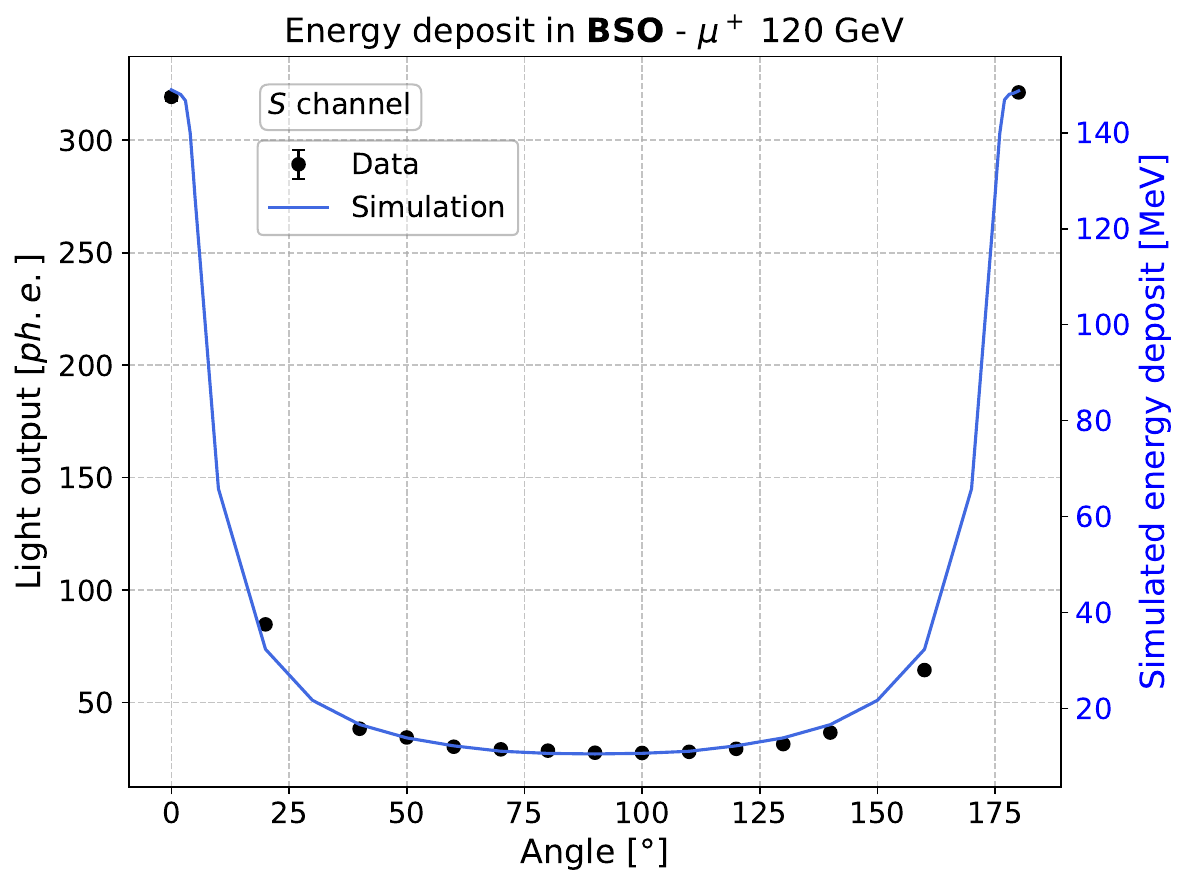}
    \end{subfigure}
    \caption{Photoelectron yields measured on the $S$ channel (black points), and simulated energy deposition (green and blue lines), shown as a function of the crystal orientation angle with respect to the beam axis in $\mu^+$ beam runs.}
    \label{fig:U_plot}
\end{figure*}

Figure~\ref{fig:mu_Schannel_waveforms} shows an example of waveforms corresponding to the passage of a MIP collected in BGO and BSO on the $S$ channel. The more irregular shape of the BGO waveform is a consequence of its longer scintillation decay time, which makes statistical fluctuations in the detected photoelectron rate more visible.
For each event, the waveform from the $S$ channel was integrated after the subtraction of the pedestal, defined as the mean value computed in the first 5 ns of the waveform acquisition time. 
The corresponding distributions of waveform integrals exhibit a typical Landau shape, as shown in Figure~\ref{fig:landau_fit}.
The Landau Most Probable Value (MPV), extracted from fits to the waveform integral distribution, was then converted into a photoelectron yield using the SiPM charge calibration described in Section~\ref{sec:Calibration}, and corrected for the contribution of the SiPM dark count rate. Figure~\ref{fig:U_plot} shows the MPV photoelectron yield as a function of the beam crossing angle for both BGO and BSO crystals.
The MPVs of the simulated energy deposit distributions, fitted with the same Landau model, are shown in the same plots with a different vertical axis, chosen for visualization purposes.
The simulation reasonably reproduces both the photoelectron yield trend and the peak-to-valley variation of a factor of $\sim$13, which is consistent with expectations based on the crystal geometry, \emph{i.e.} the ratio between the crystal longitudinal and transverse size. Small discrepancies between data and simulation are observed and are attributed to a few degrees of misalignment in the experimental setup, which was estimated by taking the average of the angular shift necessary to match the two distributions. This misalignment is quantified as  $(-2.1 \pm 2.7)^\circ$ for BGO and $(-1.5 \pm 3.0)^\circ$ for BSO, and a corresponding angular correction is applied throughout the rest of the analysis to account for this bias.

\begin{figure}[htp!]
    \centering
    \includegraphics[width=0.99\linewidth]{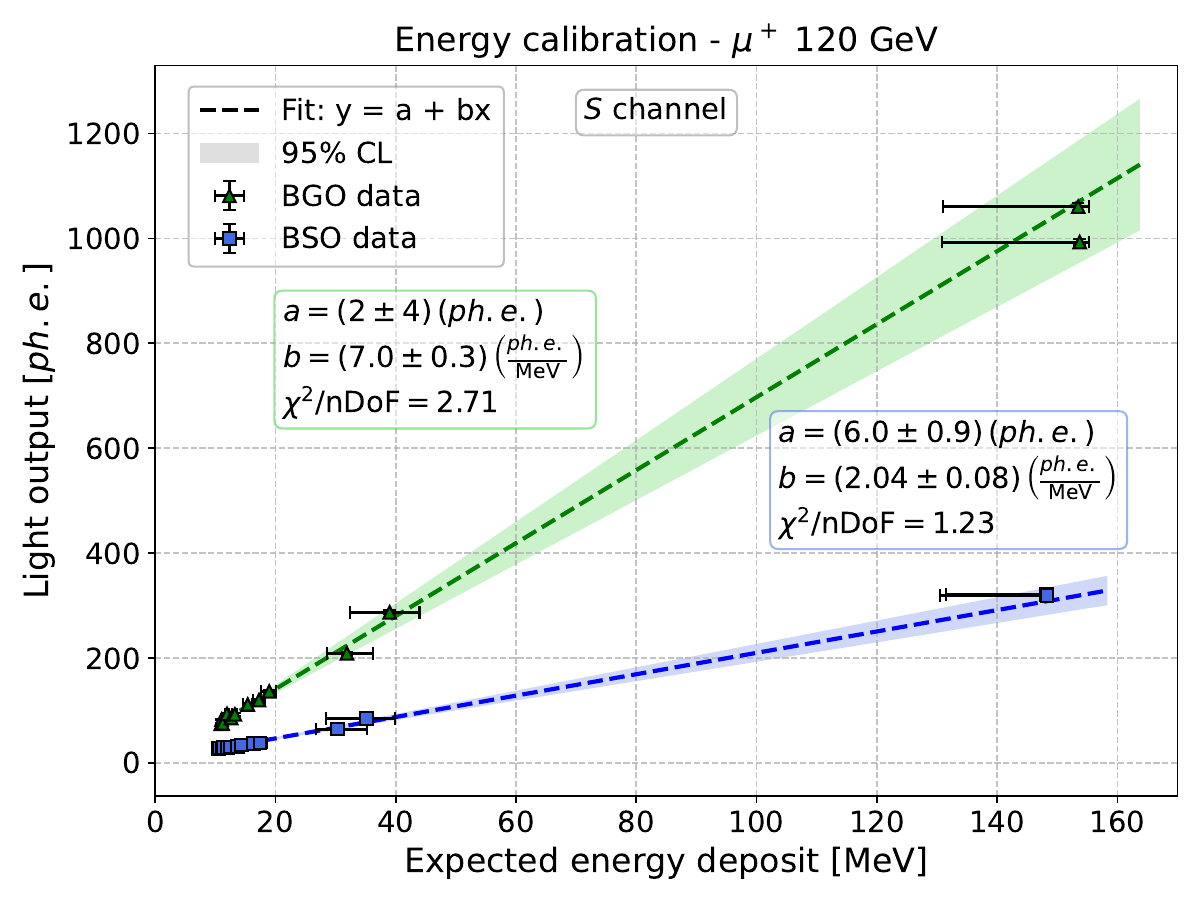}
    \caption{Photoelectron yield measured on the $S$ channel as a function of the expected deposited energy in $\mu^+$ beam runs. Experimental data points for BGO (green) and BSO (blue) are shown with error bars. Dashed lines indicate linear fits of the form $y = a + b x$, and the shaded bands represent the 95\% confidence intervals.}
    \label{fig:E_calibration}
\end{figure}

Assuming proportionality with deposited energy and neglecting residual angular effects, data and simulation were matched point-by-point at corresponding angles, as shown in Figure~\ref{fig:E_calibration}. The error bars in the plot include the uncertainty associated with the offset angle estimation. 
The asymmetric horizontal error bars at the highest deposited energies originate from the uncertainty on the crystal angular alignment near the beam-parallel configurations ($0^\circ$ and $180^\circ$): since the nominal deposited energy is maximal in these configurations, any angular deviation can only reduce the deposited energy, resulting in an asymmetric uncertainty toward lower deposited energy values.
A strong linear correlation, with a Pearson coefficient $r>0.99$ for both crystals, is observed, demonstrating good agreement between measured light output and deposited energy over the explored range. A linear fit provides the $S$ channel energy calibration factor for both BGO and BSO crystals (Figure~\ref{fig:E_calibration}). The corresponding photoelectron yields per MeV are summarized in Table~\ref{tab:phe_yields}. The uncertainties reported are obtained by propagating the fit result errors and the additional systematic uncertainty from the calibration procedure mentioned in Section~\ref{sec:Calibration}.

\begin{table}[h!]
    \centering
    \begin{tabular}{cc}
        \toprule
        \toprule
         Crystal & Light output [$ph.e.$/MeV] \\ 
         \midrule
        BGO & $7.0 \pm 0.8$ \\ 
        BSO & $2.0 \pm 0.2 $  \\ 
        \bottomrule
        \bottomrule
    \end{tabular}
    \caption{Photoelectron yield per MeV measured for BGO and BSO crystals on the S channel SiPM (Hamamatsu S14160-3010).}
    \label{tab:phe_yields}
\end{table}

\begin{figure*}[htb!]
    \centering
    \begin{subfigure}{0.485\linewidth}
        \includegraphics[width=0.99\linewidth]{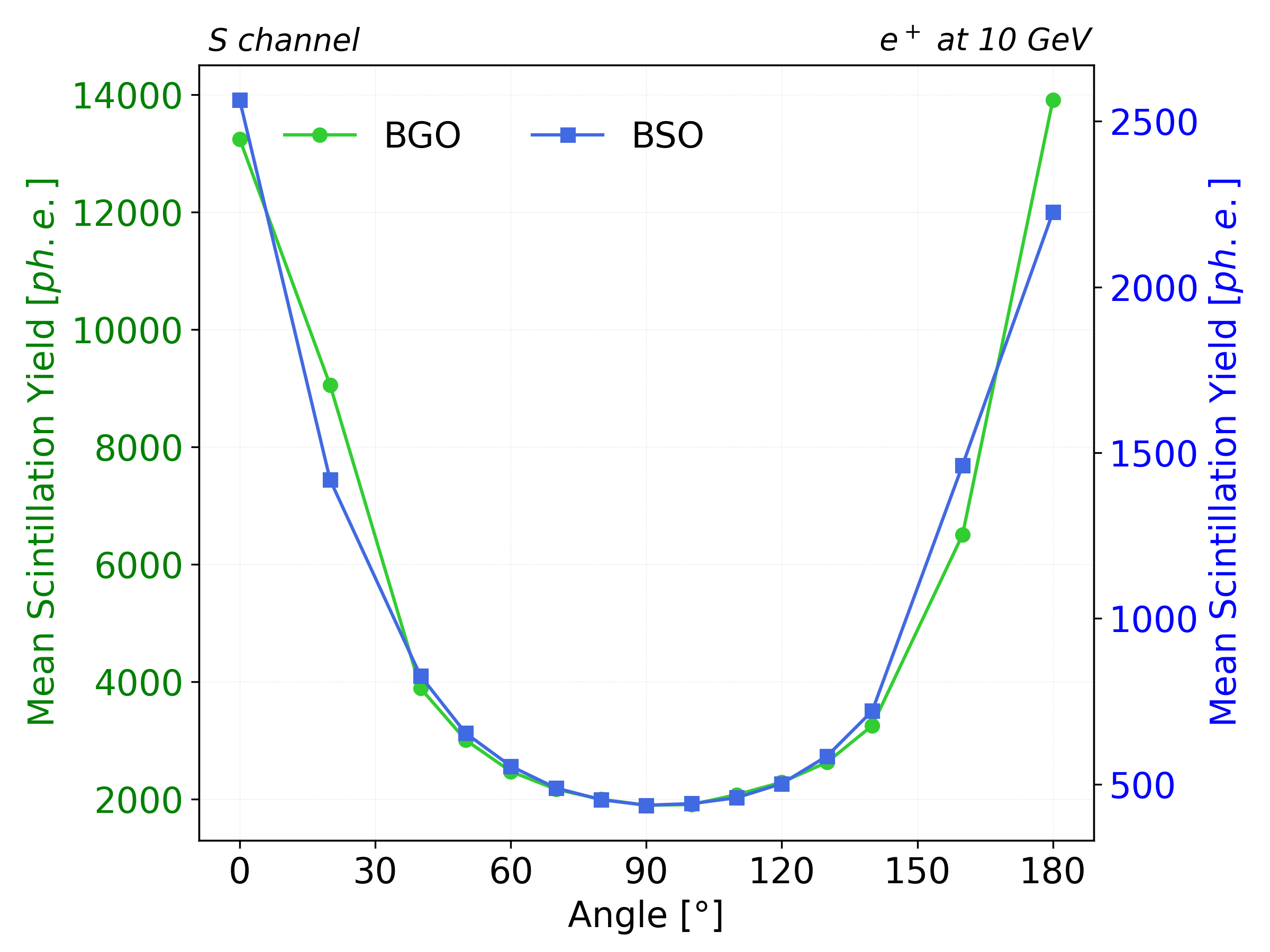}
    \end{subfigure}     
    \begin{subfigure}{0.4225\linewidth}
        \includegraphics[width=0.99\linewidth]{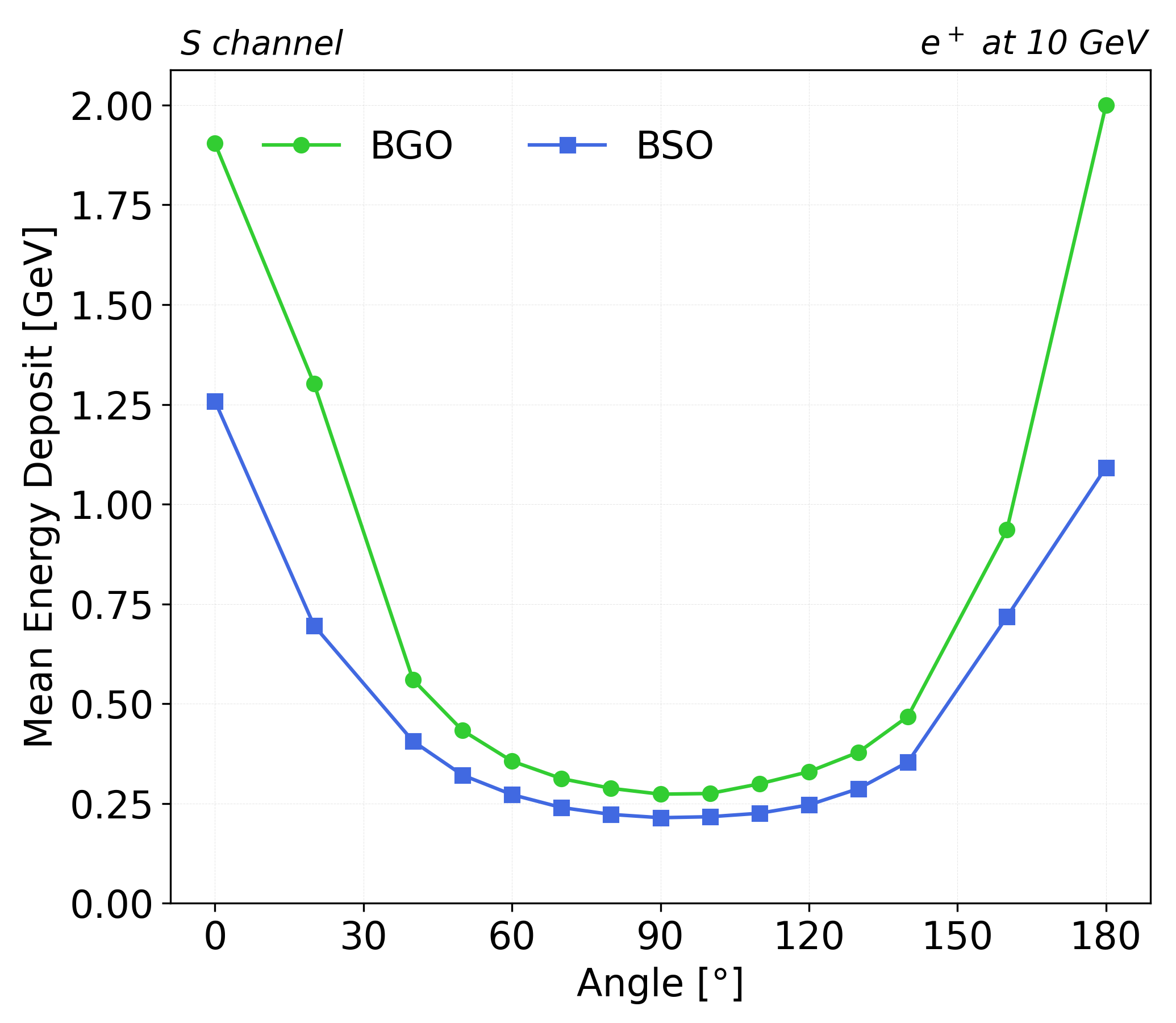}
    \end{subfigure}
    \caption{Average photoelectron yield measured on the $S$ channel (left) and average reconstructed deposited energy (right)  as a function of the crystal orientation angle in $e^+$ beam runs. Results are shown for BGO (green) and BSO (blue) crystals.}
    \label{fig:e_energy_vs_angle}
\end{figure*}

In $e^+$ runs at 10 GeV, the limited lateral dimensions of the crystal (1.2~cm), compared to the Moliere radius ($\sim$2.2 cm), resulted in a significant lateral leakage of the electromagnetic shower, reaching, on average, up to $\sim$98\% in the $90^\circ$ configuration. The event-by-event deposited energy was estimated by using the $S$ channel light output and the energy calibration factors obtained from the muon runs.
For each event, the waveform from the $S$ channel was integrated and converted into photoelectrons following the same procedure used for muons.
Figure~\ref{fig:e_energy_vs_angle} (left) shows the mean scintillation yield on the $S$ channel as a function of the crystal orientation angle for BGO and BSO.
The scintillation signal exhibits an almost symmetric angular dependence around \( 90^\circ \). For BGO, the measured yield ranges from 2000~$ph.e.$\ at \( 90^\circ \) to $\sim$13000~$ph.e.$\ at \( 0^\circ \) and \( 180^\circ \), while for BSO it varies from 500~$ph.e.$\ at \( 90^\circ \) to $\sim$2500~$ph.e.$\ at \( 0^\circ \) and \( 180^\circ \).

The scintillation photoelectron yields were then converted into deposited energies using the light output calibration factors from Table~\ref{tab:phe_yields}. The mean reconstructed energy is shown in Figure~\ref{fig:e_energy_vs_angle} (right).
A systematic difference between the measured deposited energy for the two crystals is observed, with BGO yielding values typically about 20\% higher than BSO over most of the explored angular range. This discrepancy was not expected based on the intrinsic properties of the materials and is likely attributable to experimental effects not fully controlled in the setup, including variations in light collection efficiency arising from differences in optical coupling between the crystal and the SiPM, angular bias, and uncertainties in the photo-response calibration associated with the different amplification settings for BGO and BSO. 
The observed 20\% difference is conservatively taken as the systematic uncertainty on the deposited-energy estimate obtained from the $S$ channel.

\subsection{Cherenkov yield extraction}
The Cherenkov signal in the $C$ channel, produced by electromagnetic showers in positron runs, is extracted through a dedicated waveform analysis. Although an optical filter was used to suppress the scintillation light, a residual contamination is expected. A pulse shape analysis is therefore employed to effectively disentangle the Cherenkov and scintillation contributions.
The separation strategy exploits the difference in the temporal distribution between the prompt Cherenkov and the slower scintillation emission mechanisms. To exploit this feature, a template-fitting procedure was applied to the digitized waveforms, enabling event-by-event decomposition of the Cherenkov and scintillation components.
For each component, a waveform template $T(t)$ is built by convolving the expected photon arrival time distribution $P(t)$ at the SiPM surface with the photodetector single-photon response function $F_\text{SPR}(t)$. 
Single-photon response profiles were obtained by interpolating the SiPM calibration waveforms, which were recorded using a pulsed laser, and normalizing them to the single-photon amplitude, as discussed in Section~\ref{sec:Calibration}.

\begin{figure}[t]
    \centering
    \includegraphics[width=.99\linewidth]{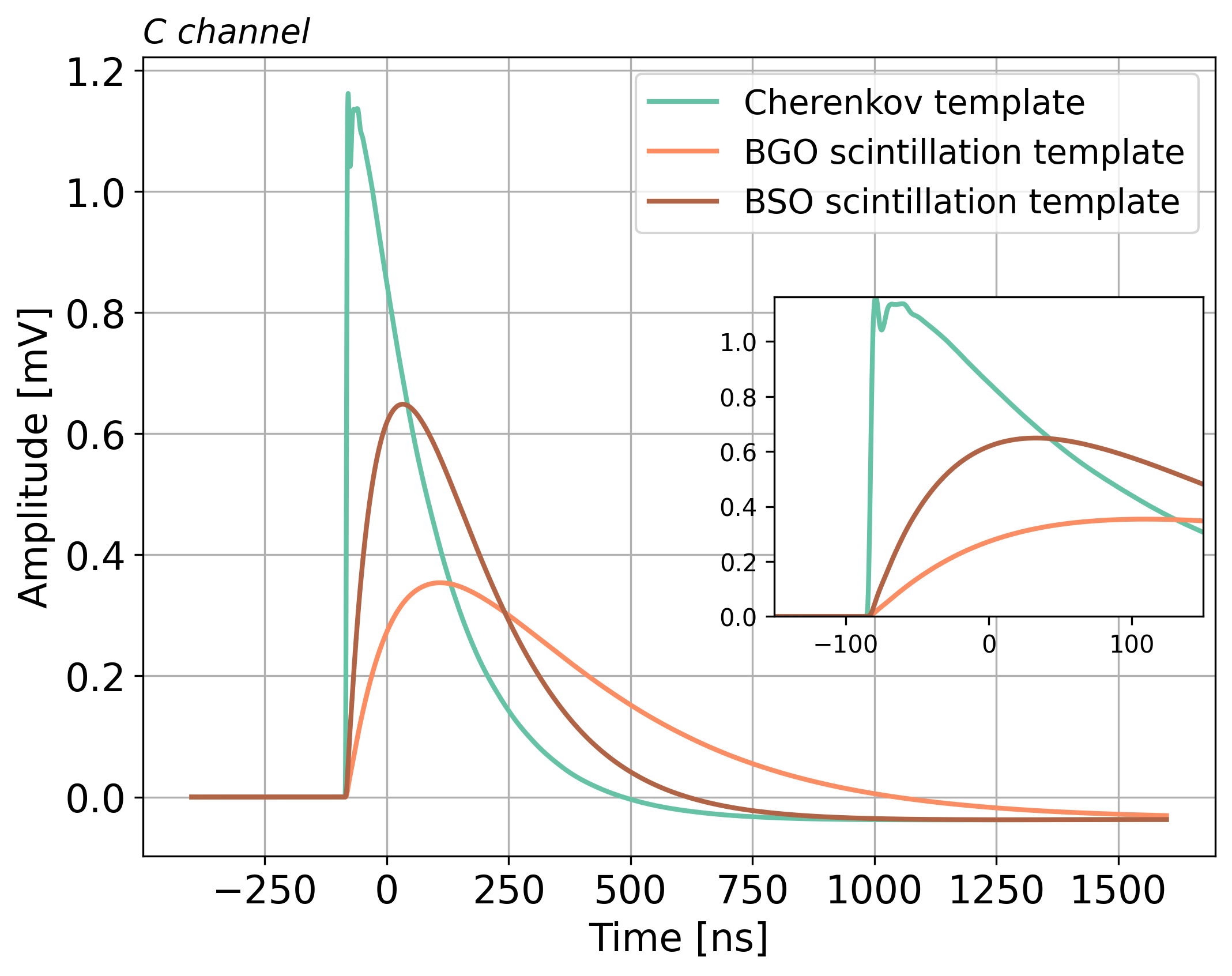}
    \caption{Templates of Cherenkov and scintillation pulse components from the $C$ channel with a 18 dB preamplifier gain in BGO and BSO. Each template is constructed by convolving the expected Cherenkov or scintillation photon arrival-time distribution at the SiPM surface with the measured single-photoelectron response of the photodetector. The inset shows a zoom on the the rising edges. The structure visible at the onset of the Cherenkov template is a reproducible feature of the single-photoelectron response measured in our setup and is therefore naturally included in the waveform template used for the fit. The templates are normalized to the single-photoelectron charge.}
    \label{fig:templates}
\end{figure}

The scintillation photon arrival time distribution, $P(t)$, was modeled using the crystal decay times extracted in Section~\ref{sec:crystal_samples}, while a prompt emission was assumed for the Cherenkov component. The function used to fit the measured waveforms is given by
\begin{equation}
f(c,s,t_0) = c\cdot T_c(t - t_0) + s \cdot T_s(t - t_0)
\end{equation}
where $T_c$ and $T_s$ are the Cherenkov and scintillation waveform templates, respectively; $c$ and $s$ represent the fitted numbers of detected Cherenkov and scintillation photoelectrons, respectively; and $t_0$ is the signal pulse arrival time. The parameters $c$, $s$, and $t_0$ are treated as free parameters in the fit, which is performed via a $\chi^2$ minimization.
Figure~\ref{fig:templates} shows the $T_c$ and $T_s$ templates used in the fitting procedure to extract the individual photon contributions from the $C$ channel SiPM. 

The fit procedure has been used to extract the yield of Cherenkov and scintillation photoelectrons event-by-event for each of the runs performed with positrons.
Pulses with a good fit quality were selected by requiring consistency between the integral of the waveform data and the integral of the fitted waveform within 25\%.
An additional requirement was imposed on the pulse amplitude, requiring it to be larger than 5 times the pedestal root mean square. Overall, these selections retained typically more than 80\% of the events.

\begin{figure*}[tbp]
    \centering
    \begin{subfigure}{0.45\linewidth}
        \includegraphics[width=\linewidth]{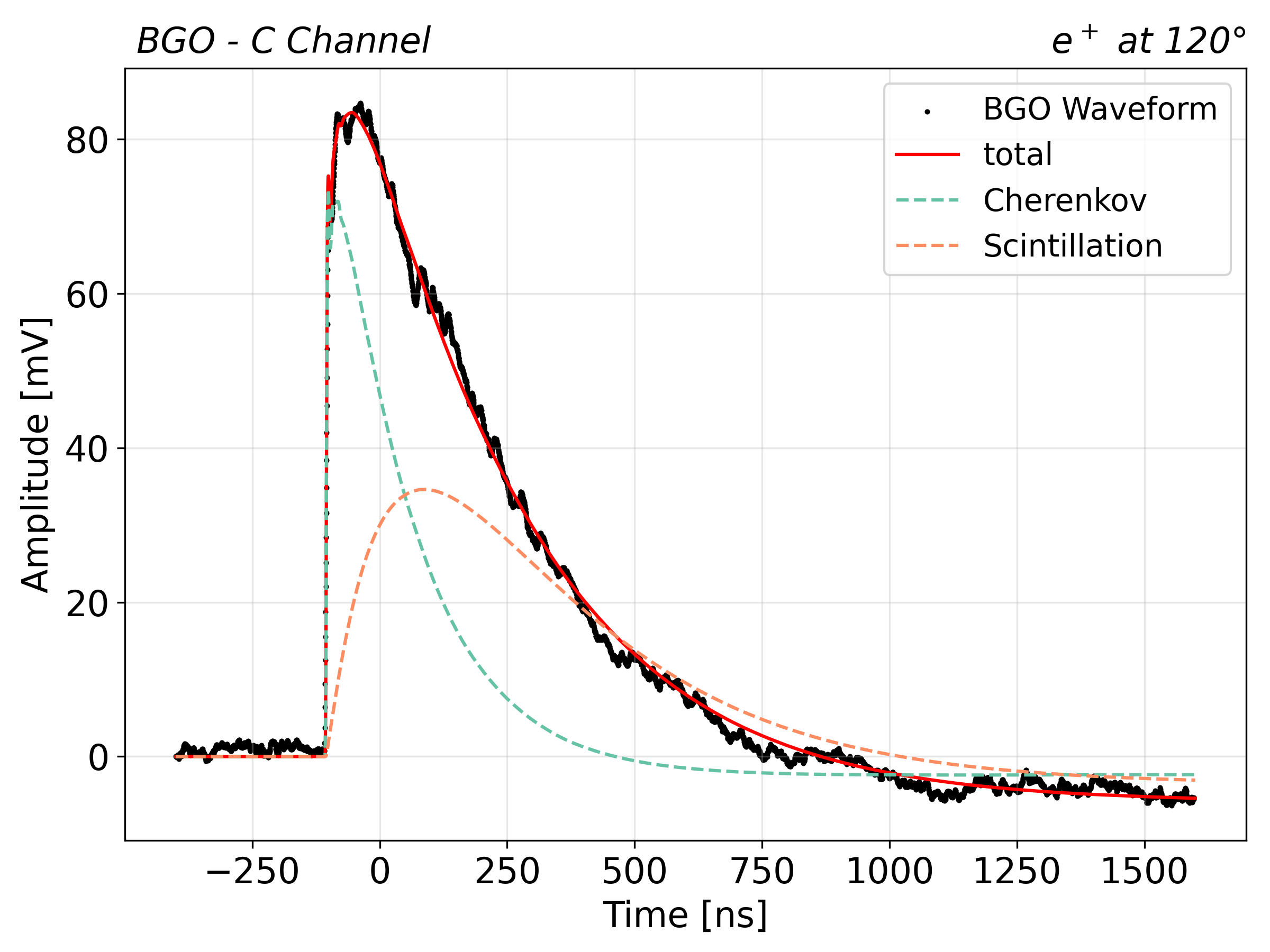}
        \caption{BGO $C$ channel at 120$^\circ$}
    \end{subfigure}
    \begin{subfigure}{0.45\linewidth}
        \includegraphics[width=\linewidth]{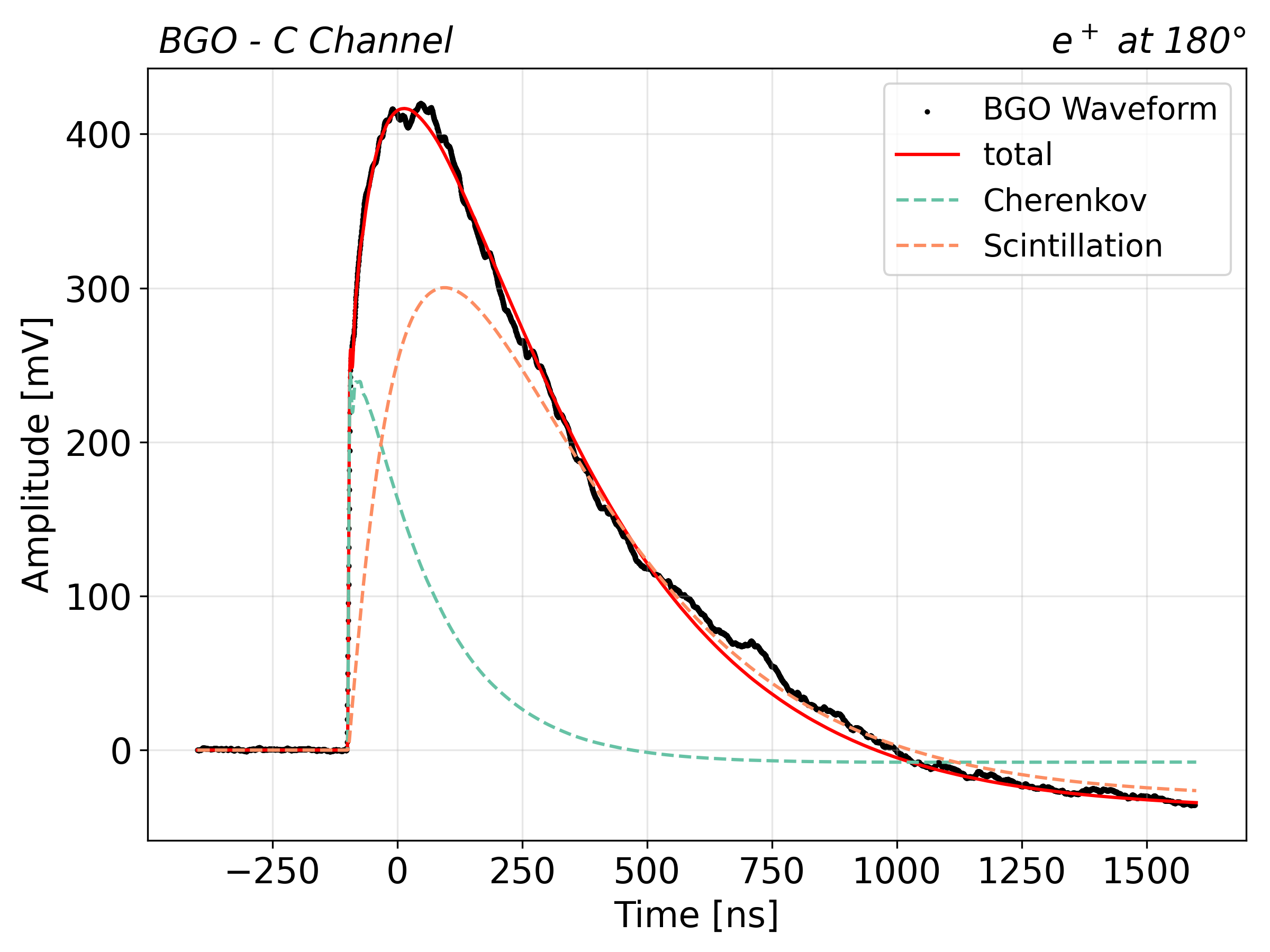}
        \caption{BGO $C$ channel at 180$^\circ$}
    \end{subfigure}\\
    \begin{subfigure}{0.45\linewidth}
        \includegraphics[width=\linewidth]{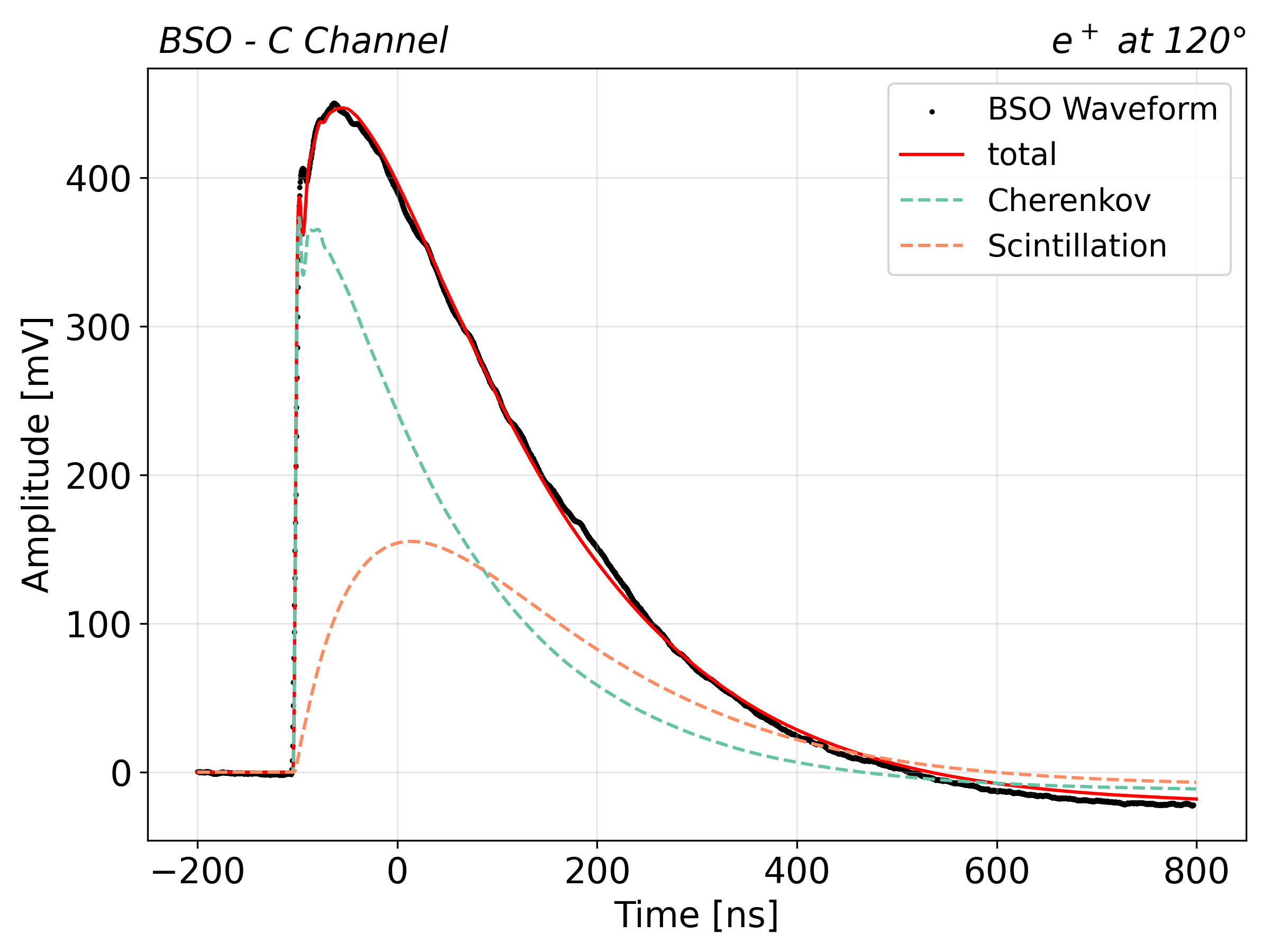}
        \caption{BSO $C$ channel at 120$^\circ$}
    \end{subfigure}
    \begin{subfigure}{0.45\linewidth}
        \includegraphics[width=\linewidth]{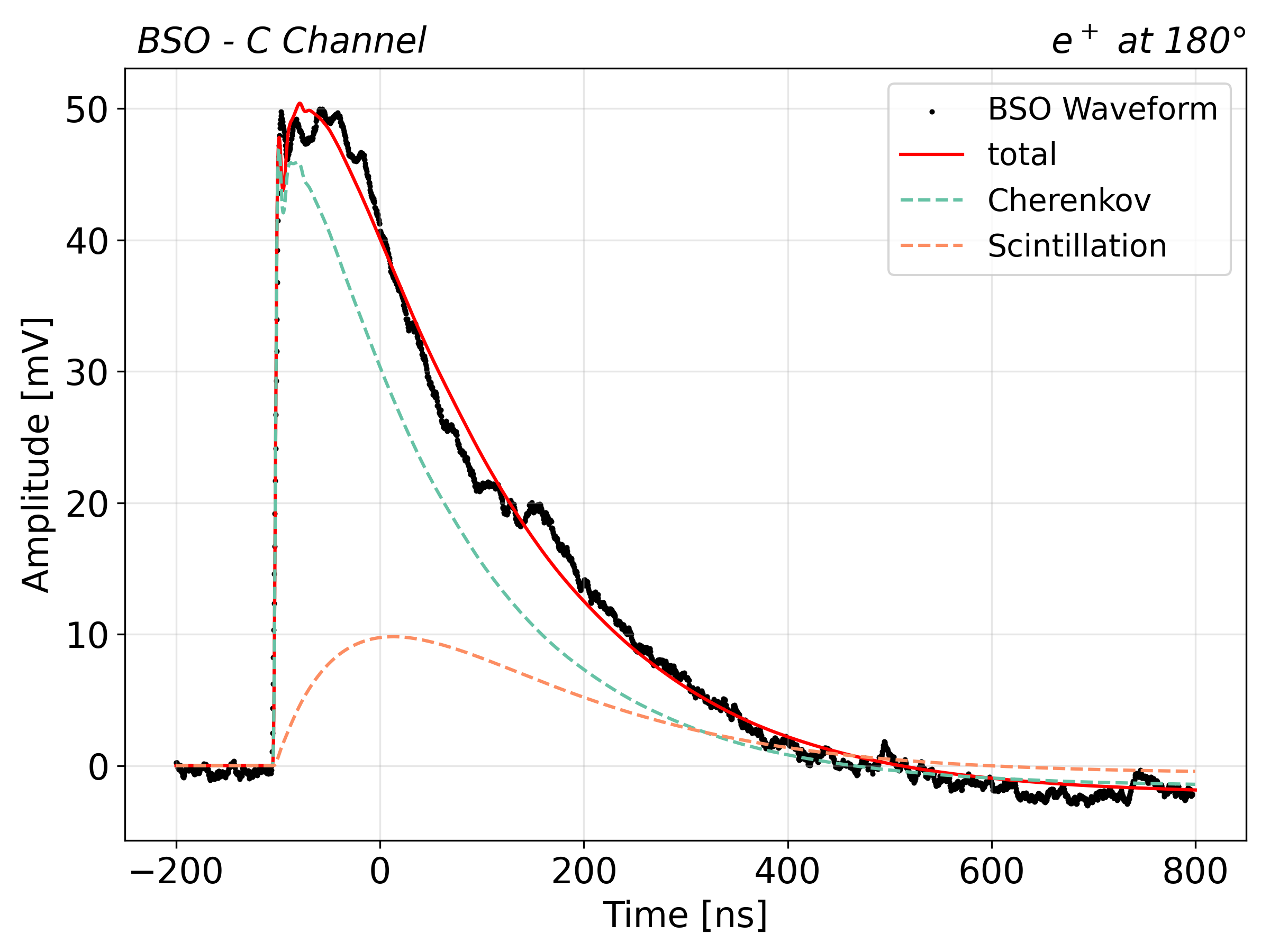}
        \caption{BSO $C$ channel at 180$^\circ$}
    \end{subfigure}

    \caption{Examples of fit to the waveforms on the $C$ channel collected in $e^+$ angular scan runs with BGO and BSO.}
    \label{fig:C-SiPM-waveforms}
\end{figure*}

Examples of fit to the waveforms in BGO and BSO crystals are shown in Figure~\ref{fig:C-SiPM-waveforms}, for two representative beam angle configurations ($120^\circ$ and $180^\circ$). 
The fit disentangles the prompt Cherenkov component from the slower scintillation contribution, with the fast leading edge well described by the Cherenkov template and the longer decay tail accounted for by the scintillation component.
The model reproduces the main features of the signal for both crystals, with small residual deviations due to statistical fluctuations in the number of photons detected and a minor contribution from SiPM dark noise.

The signal shape and the relative weights of the two fitted components vary with the angle, as shown in Figure~\ref{fig:C-SiPM-waveforms}, and further discussed below in the text.

\begin{figure*}[tbp]
    \centering
    \begin{subfigure}{0.45\linewidth}
        \includegraphics[width=\linewidth]{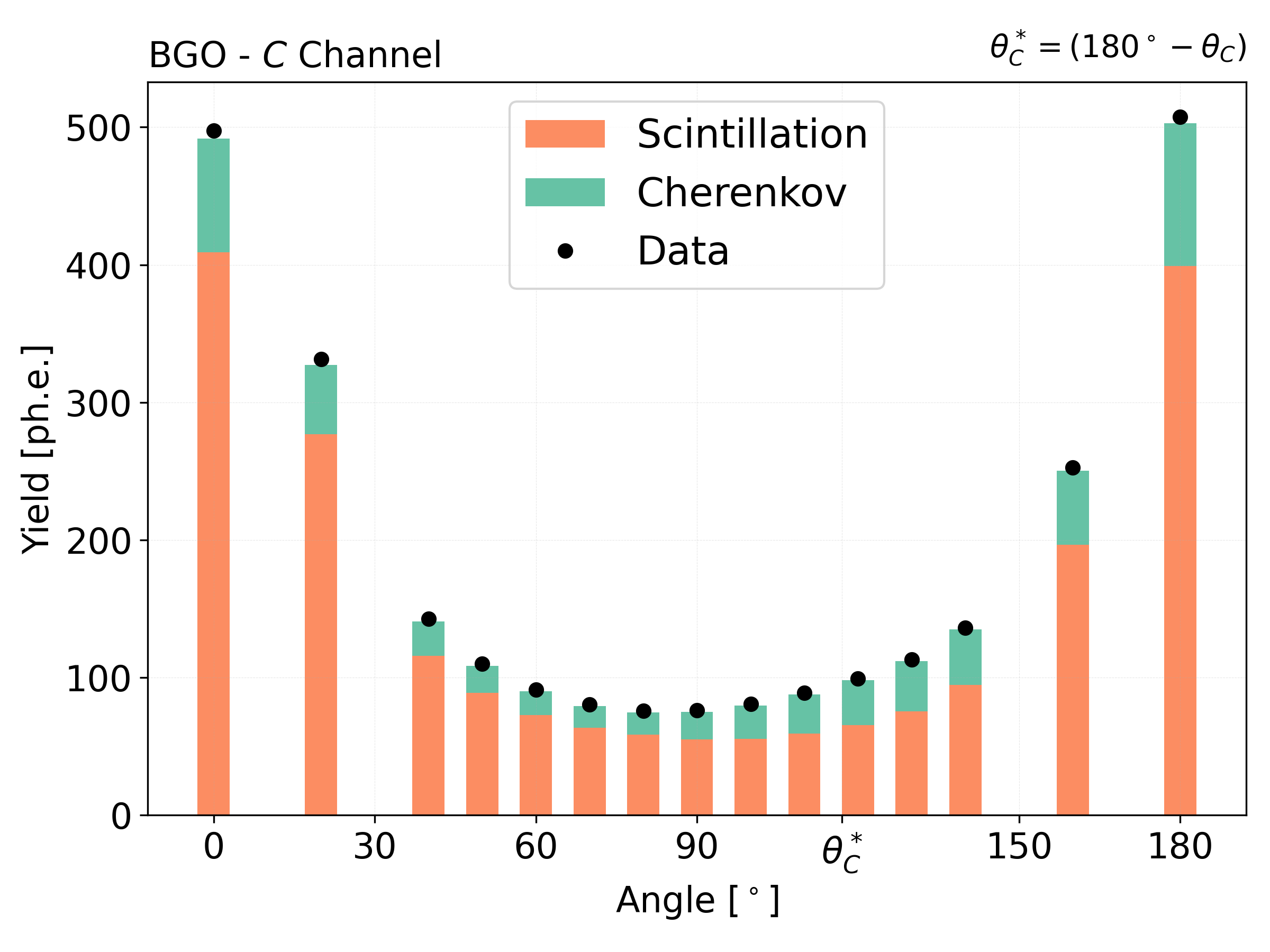}
        \caption{BGO crystal distribution}
        \label{fig:stacked_plot_BGO}
    \end{subfigure}
    \begin{subfigure}{0.45\linewidth}
        \includegraphics[width=\linewidth]{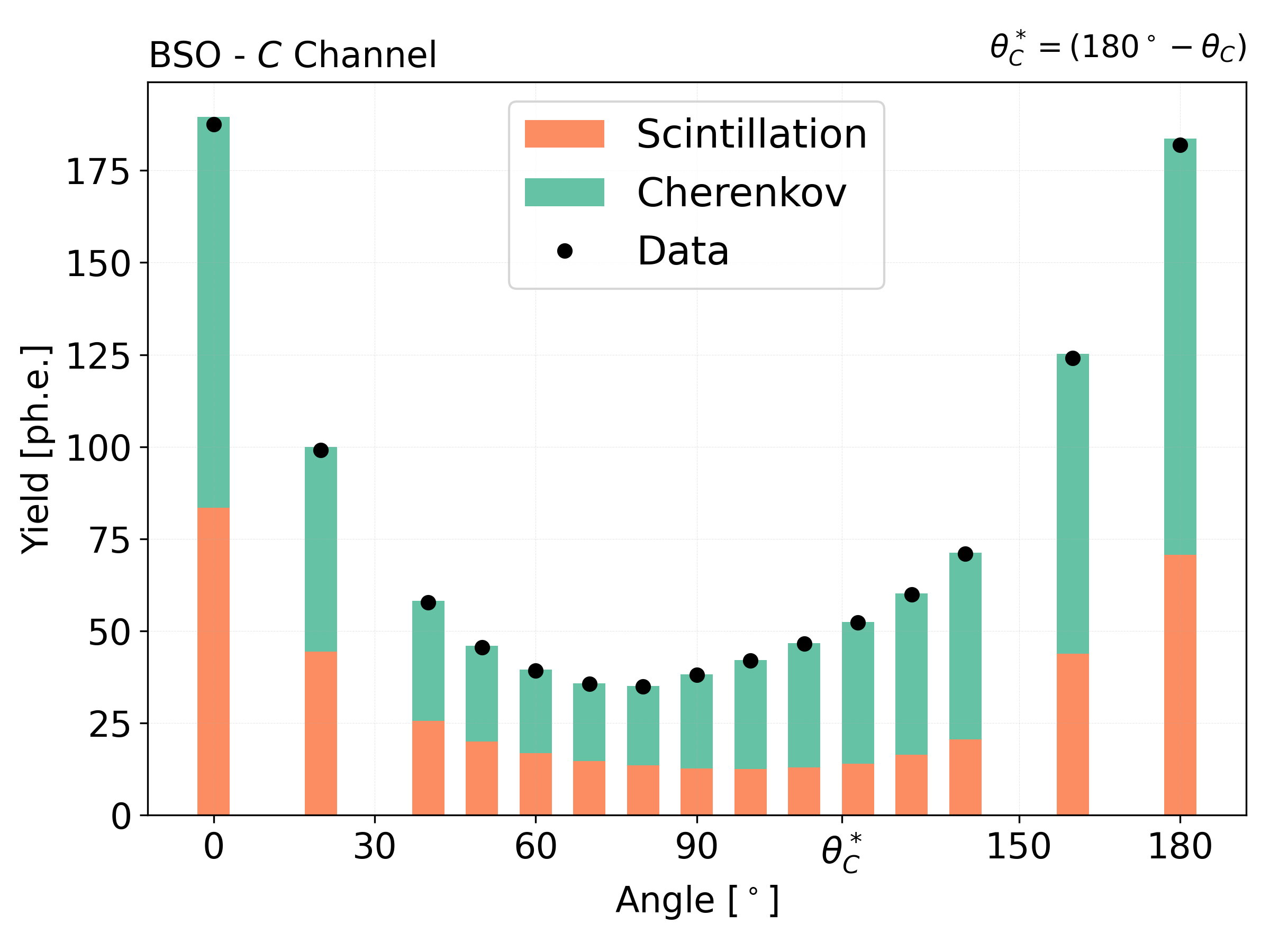}
        \caption{BSO crystal distribution}
        \label{fig:stacked_plot_BSO}
    \end{subfigure}
    \caption{Stacked distributions of scintillation and Cherenkov photoelectron yields extracted from the template-fit analysis on the $C$ channel as a function of the incidence angle for BGO and BSO crystals.}
    \label{fig:fit_results_stacked}
\end{figure*}

\begin{figure*}[htbp]
    \centering
    \begin{subfigure}{0.45\linewidth}
        \includegraphics[width=\linewidth]{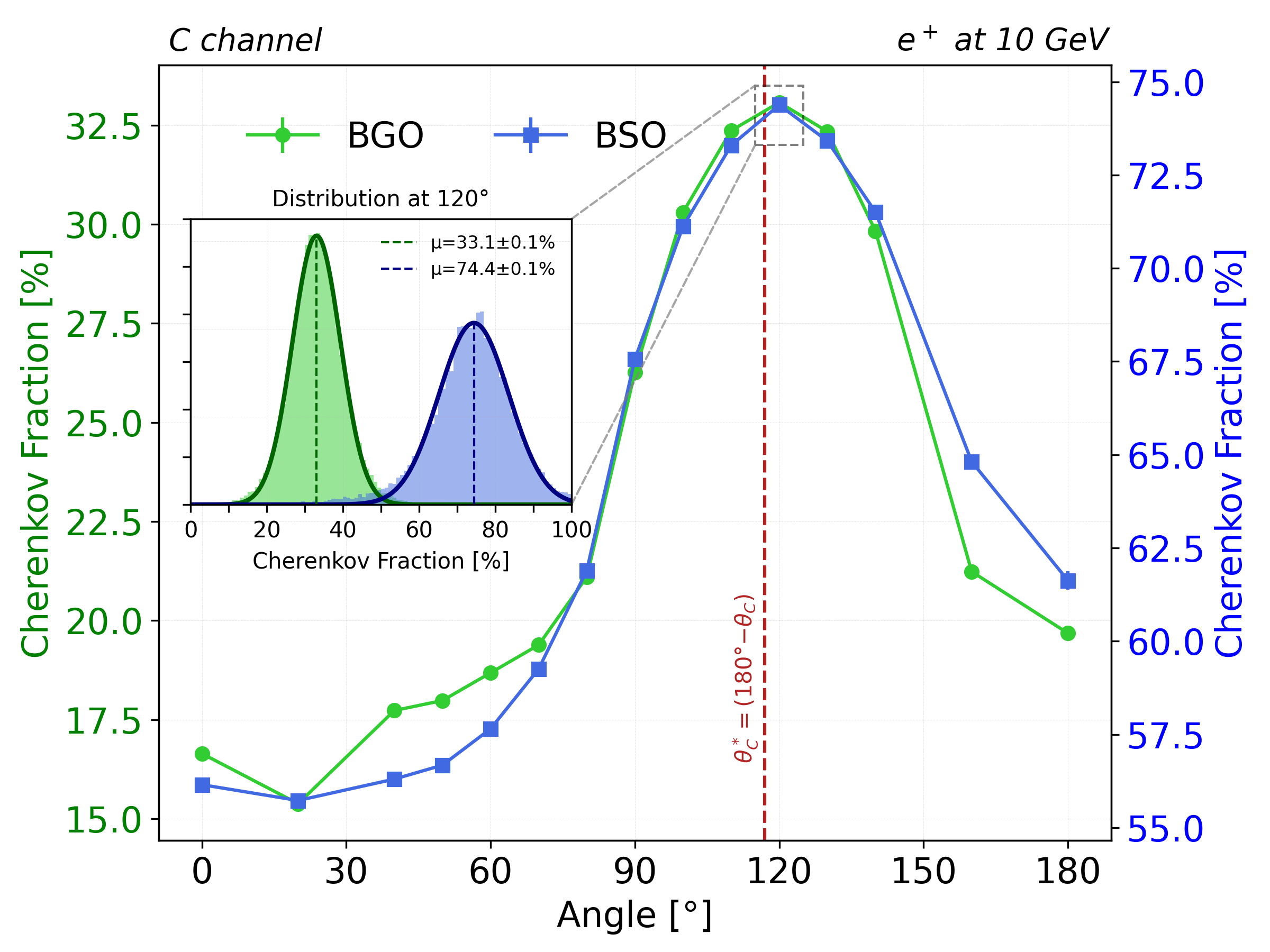}
        \caption{Cherenkov fraction in the $C$ channel.}
        \label{fig:c_over_sipm_c}
    \end{subfigure}
    \begin{subfigure}{0.45\linewidth}
        \includegraphics[width=\linewidth]{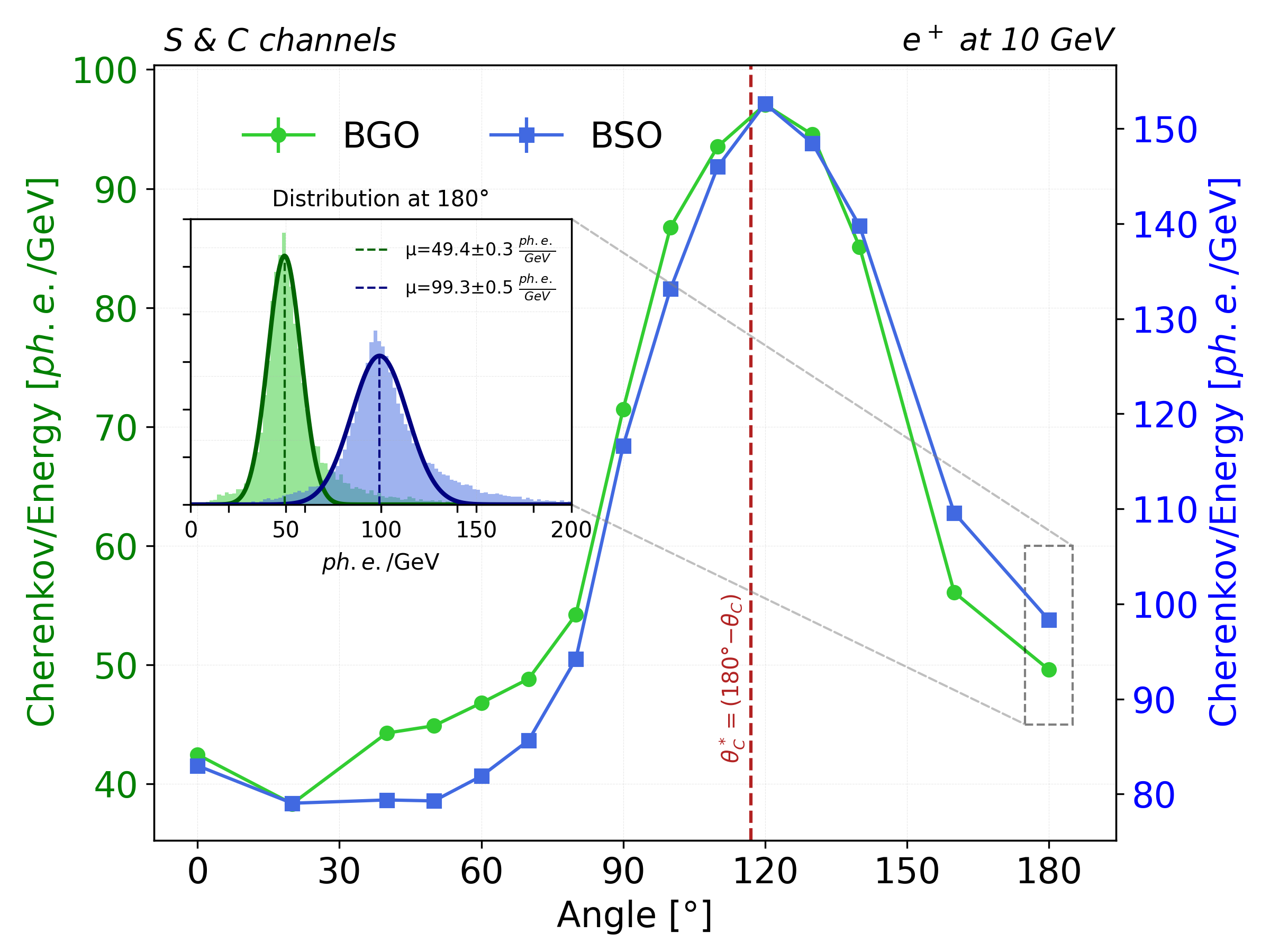}
        \caption{Cherenkov yield per energy deposit.}
        \label{fig:c_over_e}
    \end{subfigure}
     \caption{Mean values of the Cherenkov-signal distributions obtained from the template-fit analysis of the $C$ channel as a function of the incidence angle for $e^+$ runs with BGO and BSO crystals. The red dashed line indicates the expected angle $\theta_C^*=180^\circ-\theta_C$ at which the Cherenkov cone intercepts the $C$ channel. The insets show the corresponding distributions for a representative angular configuration for both crystals.}    \label{fig:fit_results}
\end{figure*}

The results of the fit over the full angular scan are summarized in 
Figure~\ref{fig:fit_results_stacked}. It shows the photoelectron yields measured in the $C$ channel as a function of the beam angle. The stacked bars represent the mean scintillation and Cherenkov contributions obtained from the template fit, while the markers indicate the photoelectron yield obtained by integrating the pedestal-subtracted waveform and using the calibration factors derived in Section~\ref{sec:Calibration}.
The sum of the two fit components provides a good description of the observed photoelectron yield across the full angular range.
The scintillation component exhibits an almost symmetric angular dependence around 90$^\circ$, similarly to the case of the $S$ channel (Figure~\ref{fig:e_energy_vs_angle}). Its yield is, however, approximately a factor of 30 lower at corresponding angles, despite the larger SiPM area and higher PDE. This suppression is due to the presence of the optical filter.
The Cherenkov contribution, by contrast, exhibits an asymmetric distribution, with a clear enhancement for angles larger than $90^\circ$, \emph{i.e.}, when the Cherenkov cone is directed toward the $C$ channel, suggesting that the fit reasonably reproduces the relative scintillation and Cherenkov contributions at different angles.
Figure~\ref{fig:c_over_sipm_c} shows the mean values of the distribution of the fitted Cherenkov fraction $C/(C+S)$ as a function of the beam crossing angles. A clear modulation is observed, reaching its maximum at $120^\circ$. This behavior is expected, since the angle at which the Cherenkov cone intercepts the $C$ channel is $\theta_C^* = 180^\circ-\theta_C = 118^\circ \,(119^\circ)$ for BGO (BSO). For simplicity in the plotting, the same Cherenkov angle is assumed for both crystals, although their actual values differ slightly (see Table~\ref{tab:crystal_param}). In the BGO crystal, a maximum Cherenkov fraction of 33\% is observed at 120$^\circ$, whereas a more favorable value is found for BSO, reaching up to $\sim$70\% at the Cherenkov angle.
This behavior reflects the intrinsically lower scintillation yield of BSO compared to BGO.

Using the event-by-event energy estimate derived from the scintillation yield measured in the $S$ channel (Section~\ref{sec:results_schannel}), a Cherenkov yield per GeV is extracted. Figure~\ref{fig:c_over_e} shows the mean values for the different beam crossing angles. 
Mean Cherenkov yields in the ranges of 100–150~$ph.e.$/GeV and 50–100~$ph.e.$/GeV are observed at $120^\circ$ and $180^\circ$, respectively. 
The corresponding total systematic uncertainty is estimated to be 23\%, obtained by combining in quadrature a 20\% uncertainty on the deposited-energy estimation with an additional 12\% uncertainty from the SiPM photo-response calibration.

The results obtained at 120$^\circ$ (the closest experimental point to the expected Cherenkov angle $\theta_C^*$) and $180^\circ$ (crystal axis aligned with the beam direction and the $C$ channel located downstream) are summarized in Table~\ref{tab:fit_results}. 

\begin{table}[h]
\centering
\caption{Cherenkov fraction and Cherenkov photoelectron yield measured for BGO and BSO crystals with the $C$ channel SiPM (Hamamatsu S14160-6050) at $120^\circ$ (closest experimental point to the expected Cherenkov angle $\theta_C^*$) and $180^\circ$ (crystal axis aligned with the beam direction, $C$ channel located downstream). The Cherenkov yield is affected by an overall uncertainty of 23\%}.
\begin{tabular}{l|cc|cc}
\toprule
\toprule

\multirow{2}{*}{Crystal} & \multicolumn{2}{c|}{$120^\circ$} & \multicolumn{2}{c}{$180^\circ$} \\
 & $\dfrac{C}{(S+C)}$ [\%] & $C\left[\dfrac{ph.e.}{\text{GeV}}\right]$ & $\dfrac{C}{(S+C)}$ [\%] & $C\left[\dfrac{ph.e.}{\text{GeV}}\right]$ \\
\midrule
BGO & 33 & 97 & 20 & 49 \\
BSO & 74 & 152 & 62 & 99 \\
\bottomrule \bottomrule
\end{tabular}
\label{tab:fit_results}
\end{table}

\section{Conclusions}
\label{sec:Conclusions}

This work presents an experimental proof-of-principle of the event-by-event separation of Cherenkov and scintillation light in high-density BGO and BSO crystals for dual-readout calorimetry. 
The separation is achieved through the use of wavelength filtering and waveform template fitting, exploiting the different spectral and temporal characteristics of the two light components. Prototypes using optical filters and SiPM readout were tested with high-energy muon and positron beams at the CERN SPS North Area. 

Scintillation yields of~$\sim$7000~and~$\sim$2000~\textit{ph.e.}/GeV are measured for BGO and BSO, respectively, using a 3$\times$3~mm$^2$ SiPM. At the same time, the analysis demonstrates that the Cherenkov component can be reliably extracted from the waveforms recorded with a 6$\times$6~mm$^2$ SiPM. The Cherenkov signal exhibits a strong angular dependence, with a maximum around~120$^\circ$, in agreement with expectations based on the Cherenkov emission angle. The Cherenkov fraction reaches approximately~33\% in BGO and~70\% in BSO, reflecting the lower intrinsic scintillation yield of the latter material. The Cherenkov yield normalized to the deposited energy lies in the range of 100–150~\textit{ph.e.}/GeV near the optimal angular configuration and 50–100~\textit{ph.e.}/GeV for parallel beam incidence, with an overall uncertainty of about 23\%, due to systematic effects. Although the tested BSO crystal yields a higher absolute number of Cherenkov photoelectrons than the tested BGO crystal, the present measurements do not establish whether this difference reflects an intrinsic property of the two materials, since the comparison is based on a single crystal sample for each material and may be affected by sample-dependent optical properties and residual experimental systematic effects.

These results validate the adopted approach for the implementation of dual-readout in homogeneous calorimeters. The measured Cherenkov yields exceed the requirement of~$\sim$50~\textit{ph.e.}/GeV for achieving an optimal stochastic term in hadronic jet energy resolution~\cite{Lucchini:2022vss,Lucchini:2020bac,Lucchini:2022goz}, while the scintillation yields are consistent with typical expectations for high-resolution electromagnetic calorimetry.
Both BGO and BSO emerge as viable candidates, each with distinct advantages: BGO offers the benefit of well-established large-scale production and a high light yield, whereas BSO provides a faster scintillation response, which is beneficial for reducing detector occupancy in high-rate environments. These complementary properties indicate that material optimization is key, and the positive findings reported in this paper for both BGO and BSO support and encourage exploration of intermediate BGSO compositions~\cite{CALA2022166527}, which potentially represent a promising path toward an optimal balance between light yield, timing performance, and Cherenkov signal purity.

This study provides a basis for the development of compact, high-performance calorimeters  for future Higgs factories, as foreseen in the IDEA detector concept~\cite{IDEAStudyGroup:2025gbt}.


\section*{Acknowledgments}

The authors would like to thank Giuseppe Passeggio (INFN Napoli) and Dr. Alcide Bertocco (INFN Napoli) for their support in the design and assembly of the mechanical setup, and Paolo Di Meo (INFN Napoli) for his contribution to the design and implementation of the front-end electronic components used in this work.

This work was carried out within the framework of the DRD Calo collaboration. It received funding from the European Union through the Next Generation EU program (Mission 4, Component 1, CUP H53D23001120006) and from the Horizon Europe ERA Widening Project TWISMA (Grant Agreement No. 101078960).
  

\bibliographystyle{unsrt}
\bibliography{bibliography}

\end{document}